\documentclass[a4paper]{article} 	
\usepackage{geometry}  		
\usepackage{graphicx,subcaption}					
\usepackage{amssymb}
\usepackage{indentfirst}
\usepackage{amsmath}
\usepackage{amsthm}
\usepackage{bm}
\usepackage{lineno}
\usepackage{setspace}
\usepackage{booktabs,multirow,array}
\usepackage{authblk}
\usepackage{graphicx}
\usepackage{float}
\usepackage[flushleft]{threeparttable} 
\usepackage{xspace}

\RequirePackage[colorlinks,citecolor=blue,urlcolor=blue]{hyperref}

\newcommand{\R}{\texttt{R}}
\newcommand{\asreml}{\texttt{ASReml-R}}

\newcommand{\TPE}{T\ensuremath{\times}PE\xspace}

\usepackage[url=false, isbn=false, eprint=false, backref=false, style= authoryear, backend=biber, maxcitenames=2, giveninits=true, maxbibnames=100, uniquename=init]{biblatex}   
\DeclareNameAlias{sortname}{family-given}
\preto\fullcite{\AtNextCite{\defcounter{maxnames}{99}}}

\DeclareCiteCommand{\cite}
  {\usebibmacro{prenote}}
  {\usebibmacro{citeindex}%
   \printtext[bibhyperref]{\usebibmacro{cite}}}
  {\multicitedelim}
  {\usebibmacro{postnote}}

\DeclareCiteCommand*{\cite}
  {\usebibmacro{prenote}}
  {\usebibmacro{citeindex}%
   \printtext[bibhyperref]{\usebibmacro{citeyear}}}
  {\multicitedelim}
  {\usebibmacro{postnote}}

\DeclareCiteCommand{\parencite}[\mkbibparens]
  {\usebibmacro{prenote}}
  {\usebibmacro{citeindex}%
    \printtext[bibhyperref]{\usebibmacro{cite}}}
  {\multicitedelim}
  {\usebibmacro{postnote}}

\DeclareCiteCommand*{\parencite}[\mkbibparens]
  {\usebibmacro{prenote}}
  {\usebibmacro{citeindex}%
    \printtext[bibhyperref]{\usebibmacro{citeyear}}}
  {\multicitedelim}
  {\usebibmacro{postnote}}
  
\DeclareCiteCommand{\textcite}
 {\boolfalse{cbx:parens}}
  {\usebibmacro{citeindex}%
    \printtext[bibhyperref]{\printnames{labelname}%
      \printtext{ (\printfield{year}\printtext{)}}}}
 {\ifbool{cbx:parens}
  {\bibcloseparen\global\boolfalse{cbx:parens}}
  {}%
 \multicitedelim}
{\usebibmacro{textcite:postnote}}

\title{Statistical Elements of a Successful Farm-scale Trial}

\author[1]{Zhanglong Cao \thanks{Corresponding author: zhanglong.cao@curtin.edu.au}}
\author[1]{Suman Rakshit}
\author[2]{Barbara Kachigunda}
\author[2]{Julia Easton}
\author[2]{Mark Gibberd}

\affil[1]{School of Electrical Engineering, Computing and Mathematical Sciences, Curtin University, Perth, WA, Australia}
\affil[2]{Centre for Crop and Disease Management (CCDM), Curtin University, Perth, WA, Australia}
 
\date{October 2026}

\begin{document}

\maketitle

\begin{abstract}
\noindent\textbf{Context or problem:} Growers and advisers increasingly run large strip trials on their own paddocks, but the rules used to design them come from small-plot research and have not been tested at farm scale. It is also unclear whether yield-monitor data should be analysed as individual pixels or as strip-plot means.

\noindent\textbf{Objective or research question:} This study evaluates which design elements determine what a strip trial can detect, and which analysis should be used for which kind of data.

\noindent\textbf{Methods:} Strip trials were simulated on a spatially autocorrelated field, varying trial length, replication, treatment allocation (randomised or systematic) and layout (long-strip or stacked-replicate), for a nitrogen response that was uniform or varied across 2--10 within-field pseudo-environments (PEs). Analyses of strip-plot means and of yield pixels were first checked for Type I error, and designs were then compared by the smallest response that a valid analysis detects with 80\% power. Further simulations added gaps in the data, unequal strip lengths, along-strip trends, spatially varying responses and alternative field heterogeneity structures. More than 3 million models were fitted.

\noindent\textbf{Results:} Treating yield pixels as independent replicates produced false positives in 32--40\% of null trials. Replication was the strongest design lever: the detectable response scaled with $r^{-0.60}$ to $r^{-1.02}$ but only as $L^{-0.45}$ or more slowly, and at equal plot area more, shorter replicates won all 66 comparisons. With complete data, strip-plot means with an autoregressive residual were the most precise analysis; when strips contained gaps, pixel-level spatial models with a strip effect halved the detectable response, whereas the spatial analysis of strip means became liberal. With PE-varying responses, spatial PE models detected per-PE effects 33--46\% smaller than PE means without a spatial term. Randomisation did not improve precision but kept the tests valid (systematic allocation: Type I error up to 23\%). Systematic allocation gave more accurate response maps.

\noindent\textbf{Conclusions:} What a strip trial can detect depends first on replication and second on whether the analysis matches the data. Strip-plot means suffice for complete data under a uniform response. For large trials with PEs, and whenever data are incomplete, pixel-level spatial models with a strip effect are the better default, and they also give better response maps. Pixel-level analyses require randomisation and a term for the strip.

\noindent\textbf{Implications or significance:} The results are condensed into a design table and a detectable-response chart that growers and advisers can use before a trial is sown.
\end{abstract}


\section{Introduction}

Randomised experimental designs have underpinned agronomic inference since Fisher formalised the logic of randomisation in the 1920s and 1930s \parencite{Fisher1934Statistical, Fisher1935Design, Verdooren2020History}. Random assignment of treatments to plots guards against confounding with uncontrolled field variation, so treatment estimates are unbiased whatever the spatial pattern of soil heterogeneity. The randomised complete block design (RCBD), split-plot, strip-plot and Latin square designs all rest on this principle, and they remain the standard designs of small-plot agricultural research \parencite{Petersen1994Agricultural, Hinkelmann2012Design}. Small-plot trials, however, rarely resemble a working farm. Plots are small, management is intensive and uniform, and the results must pass through several steps before a farmer can act on them \parencite{li2023economic}. This can take years, by which time varieties, input costs and climate may have changed.

On-farm experimentation (OFE) has become a major route for agronomic research, because digital technologies and precision machinery now let individual farms run large field experiments \parencite{Lacoste2022OFE}. An OFE trial sits inside the farmer's own production system. It uses commercial equipment and the farmer's own management, so its results apply directly to decisions on that farm \parencite{Cook2013Onfarm, evans2020Assessment}. Strip trials have become the most widely adopted OFE format because they fit the linear geometry of field operations: treatment strips are as wide as the application equipment and as long as the paddock, and yield data are collected during normal harvest without additional field passes \parencite{Pringle2004FieldScale, Rakshit2020Novel}. A grower or adviser planning a strip trial must decide how long the strips should be, how many replicates to lay out, whether the treatments should be randomised, and whether to run full-length strips or to stack two shorter sets of strips end to end. Each of these decisions determines how small an effect the trial can detect and whether its conclusions can be trusted, but there is little quantitative guidance on which of them matter most.

Practice rests on four design intuitions, most of them carried over from small-plot experimentation:
(i)~longer strips give more reliable results;
(ii)~more replicates give more reliable results;
(iii)~randomised treatment allocation is preferable to systematic allocation; and
(iv)~the long-strip layout, in which every strip runs the length of the paddock, is preferable to a stacked layout that fits more replicates into the same paddock.
The third has the deepest roots. Randomisation protects the validity of significance tests because it prevents treatment from being confounded with spatial trends \parencite{Piepho2013Why, Pringle2004FieldScale, Selle2019Flexible}. When the goal is a continuous map of the treatment response, systematic allocation is preferable, because every location then lies within a predictable distance of each treatment level \parencite{Rakshit2020Novel, cao2022bayesian, cao2024optimal, evans2020Assessment}. Even under randomised allocation, spatial correlation among observations can bias treatment comparisons if the model ignores it \parencite{RRICHTER2015Treatment}. For the classical goal of testing a few categorical treatment levels the evidence is thinner: \textcite{Piepho2011Statistical} discussed the trade-offs for on-farm trials, and in a simulation of field-scale designs \textcite{Alesso2019Experimental} found that randomisation increased the uncertainty of treatment estimates. The first two intuitions have older support. \textcite{Wuest1994Increasing} found that the experimental error of combine-width on-farm plots fell with plot length following a logarithmic decay, and \textcite{Alesso2019Experimental} found that treatment estimates became more accurate with more, smaller experimental units. Both agree with the empirical law of \textcite{Smith1938Empirical}, under which the variance of a plot mean falls more slowly than in proportion to plot size. None of the four intuitions, however, has been quantified on a common scale for the geometry of commercial strip trials, where strips span hundreds of metres and cross large spatial heterogeneity.

Judging a design requires an analysis, and at farm scale the choice of analysis is itself open. GPS-equipped harvesters record yield at 1--3\,m intervals, so a single strip contributes hundreds of yield pixels, often in several columns across the width of the harvester. Pixels within a strip share one treatment assignment; treating them as independent replicates inflates the effective sample size and makes tests too liberal, a form of pseudo-replication \parencite{Cordoba2024OFEMeanTest, Hurlbert1984Pseudoreplication}. Two responses to this problem are in use. One aggregates the pixels to one value per strip, or per strip and zone, and analyses the means. The other keeps every pixel and models the spatial dependence with a linear mixed model, as in the pseudo-environment approach of \textcite{stefanova2023}. Published methods range from simple analyses designed for routine use \parencite{Lawes2012Simple} to spatial analyses of yield-monitor data \parencite{Griffin2008Spatial}. \textcite{Tanaka03042026} note that the resolution at which OFE data are analysed can substantially affect the covariance model and the decisions drawn from it, but when each approach is preferable has not been established.

A second difficulty is that power, the usual design criterion, is evaluated at one assumed effect size. If that effect is large, most reasonable designs detect it almost every time, and power then says nothing about which design is better. The smallest effect a design can detect, the minimum detectable effect of evaluation research \parencite{Bloom1995Minimum}, does not have this problem. \textcite{Stroup2002Power} showed how the power of spatial mixed-model analyses can be computed from the noncentrality of the $F$-test, and used it to compare design and analysis strategies on spatially variable fields.

Paddocks are also not homogeneous in their response to treatment. Farmers and agronomists increasingly delineate pseudo-environments (PEs): spatially contiguous areas with similar soil properties, elevation or agronomic history. PEs are usually derived by clustering historical yield maps, proximal soil sensing, digital soil maps and remote-sensing indices \parencite{Heidari2026Evolution, bansod2013application, Tanaka03042026}. When the response varies across PEs, a treatment-by-PE (\TPE) interaction arises, analogous to the genotype-by-environment interaction of multi-environment trials (MET), with treatment in the role of genotype and PE in that of environment \parencite{stefanova2023}. For design, the PE adds further questions: whether a trial can support a separate response estimate in each of several PEs, what the multiplicity of PE-level tests costs, which analysis gives the most precise per-PE estimates, and how the PEs should be delineated \parencite{Piepho2011Statistical}.

Simulation studies of OFE have compared designs on economic return \parencite{li2023economic} and on the accuracy and Type I error of treatment estimates under several estimation methods \parencite{Alesso2019Experimental}, and \textcite{Alesso2021Design} showed that systematic designs with small or intermediate plots recover site-specific response maps best. \textcite{Marchant2019Establishinga} assessed the precision and robustness of farmers' crop experiments, and \textcite{Tanaka2021Assessment} compared design and analysis frameworks for on-farm experimentation, including spatial linear mixed models, through a simulation of wheat yield. Here we use a Monte Carlo simulation of farm-scale strip trials to test the four design intuitions and to identify which analysis suits which data. We extend the earlier work in three ways: trial length, replication, allocation and layout are varied jointly for long-strip and stacked strip trials; designs are scored by the minimum detectable response; and aggregated and pixel-level analyses are compared on complete and incomplete data. The study makes five contributions:
\begin{enumerate}
    \item It identifies which analyses can be used to judge a strip-trial design. Pixel-level analyses without a term for the experimental unit are invalid (Type I error 32--40\%); strip-plot means, or pixel models with a strip effect, are valid under randomisation (Section~\ref{sec:calibrate}).
    \item It adapts the minimum detectable effect to strip trials as the minimum detectable response at 80\% power, computed from the simulated error covariance of each design and analysis, and checks it against simulated power (Section~\ref{sec:detect}).
    \item It ranks the four design elements on that criterion: replication is the strongest lever and length a weaker one; long strips are at least as precise as stacked layouts plot for plot, but stacking can win on a fixed paddock by buying replicates; and randomisation buys validity rather than precision (Sections~\ref{sec:length-reps}--\ref{sec:robust}).
    \item It shows when pixel-level analyses pay off. With complete data, strip-plot means are the most precise; when strips contain gaps, pixel-level spatial models halve the detectable response, and they also give better response maps. Pixel-level models need randomisation and a strip effect to be valid (Section~\ref{sec:pixel}).
    \item For PE-varying responses, it shows that spatial PE models, and pixel-level PE models with a strip effect in particular, detect per-PE effects 33--46\% smaller than PE means without a spatial term. Per-PE power holds as PEs multiply, but multiplicity then dominates; per-PE tests must be conditional $F$-tests; and PE boundary placement matters far more than PE size (Section~\ref{sec:pe}).
\end{enumerate}
Section~\ref{sec:methods} describes the simulation and the performance measures, Section~\ref{sec:results} reports the results and combines them in a design table (Section~\ref{sec:framework}), and Section~\ref{sec:discussion} discusses their implications and limitations.

\section{Methodology}
\label{sec:methods}

\subsection{Study logic}
The objects of study are four design elements: trial length $L$, number of replicates $r$, treatment allocation (randomised or systematic) and layout (long-strip or stacked-replicate). Analysis models enter as the instrument through which a design is judged. The study proceeds in three steps. First, every analysis model is checked for Type I error, and only analyses that hold the nominal rate are used to judge designs. Second, each design is scored by the smallest response that a valid analysis detects with 80\% power. Third, the design elements are compared on that score under a uniform treatment response (Scenario~A). A response that varies across PEs (Scenario~B) is used to evaluate PE-level analyses, their precision and the cost of multiplicity. Because the best analysis may depend on the data, a set of follow-up simulations then compares aggregated and pixel-level analyses on complete and incomplete data. Two data representations are used: the validity checks and robustness runs of the main grid use the centre pixel column of each strip, and the design comparisons and follow-up runs use all three columns (Section~\ref{sec:runs}).

\subsection{Simulated field and trials}
Baseline yield was simulated using unconditional Gaussian geostatistical simulation \parencite{evans2020Assessment}:
\begin{equation*}
z(s) = \mu + \varepsilon(s)
\end{equation*}
where $\mu = 2500$\,kg\,ha$^{-1}$ and $\varepsilon(s)$ is a spatially autocorrelated zero-mean random field with an exponential variogram (partial sill $C_0 = 250{,}000$\,(kg\,ha$^{-1}$)$^2$, range parameter $a = 30$\,pixels, so that the correlation at lag $h$ is $\exp(-h/a)$). Throughout we adopt a nominal resolution of 10\,m per pixel: treatment strips ($w = 3$\,pixels) are then $\approx 30$\,m wide (about two machine widths), trial lengths of 10--110\,pixels span 100--1100\,m, and the range parameter corresponds to $\approx 300$\,m, with a practical range (correlation 0.05) of $3a \approx 900$\,m. Each 10\,m pixel is treated as a gridded yield cell. Because the results depend on the ratios of strip width and trial dimensions to the range parameter rather than on absolute distances, the nominal resolution affects only how they are expressed in metres.

The simulated field ($300 \times 800$ pixels) is not a model of a paddock but a source of background yield. In each iteration the trial footprint was placed at a uniformly random position within it, and the footprint itself plays the role of the paddock: in the long-strip layout every strip runs the full length $L$ of the simulated paddock. Random placement serves only to give each iteration a different patch of background heterogeneity. Because the field is stationary, all positions are statistically equivalent, and trial siting is not a factor of this study. All iterations drew on a single realisation of the field. Monte Carlo variation between iterations came from the trial position and from the treatment allocation, which was re-randomised in every iteration. Positions overlap and are not independent, so the reported rates are averages over positions within one realised field rather than over independent fields. Section~\ref{sec:robust} repeats the key design points on three independent realisations of the base field and of four alternative heterogeneity structures.

Three nitrogen rates $N \in \{0, 60, 120\}$\,kg\,ha$^{-1}$ were applied with response coefficient $\beta = 6$\,kg grain per kg N (true contrasts $\Delta_{60} = 360$ and $\Delta_{120} = 720$\,kg\,ha$^{-1}$, 14 and 29\% of mean yield). Scenario~A (uniform response) assumes the treatment works equally everywhere: $y(s) = z(s) + \beta N(s)$. Scenario~B (PE-varying \TPE response) partitions the trial footprint into $K$ horizontal PEs ($K \in \{2,4,6,8,10\}$ for strip trials and $K \in \{2,4,6,8\}$ for stacked trials, whose usable rows are reduced by the inter-block gap); within PE $k$ the response coefficient is $\beta_k$, evenly spaced from $0.5\beta$ to $2\beta$ (equal-weighted average $1.25\beta$). The PEs are defined by the response alone; the background yield does not differ systematically between them.

\begin{figure}[H]
    \centering
    \begin{subfigure}[t]{0.48\textwidth}
        \centering
        \includegraphics[width=\linewidth]{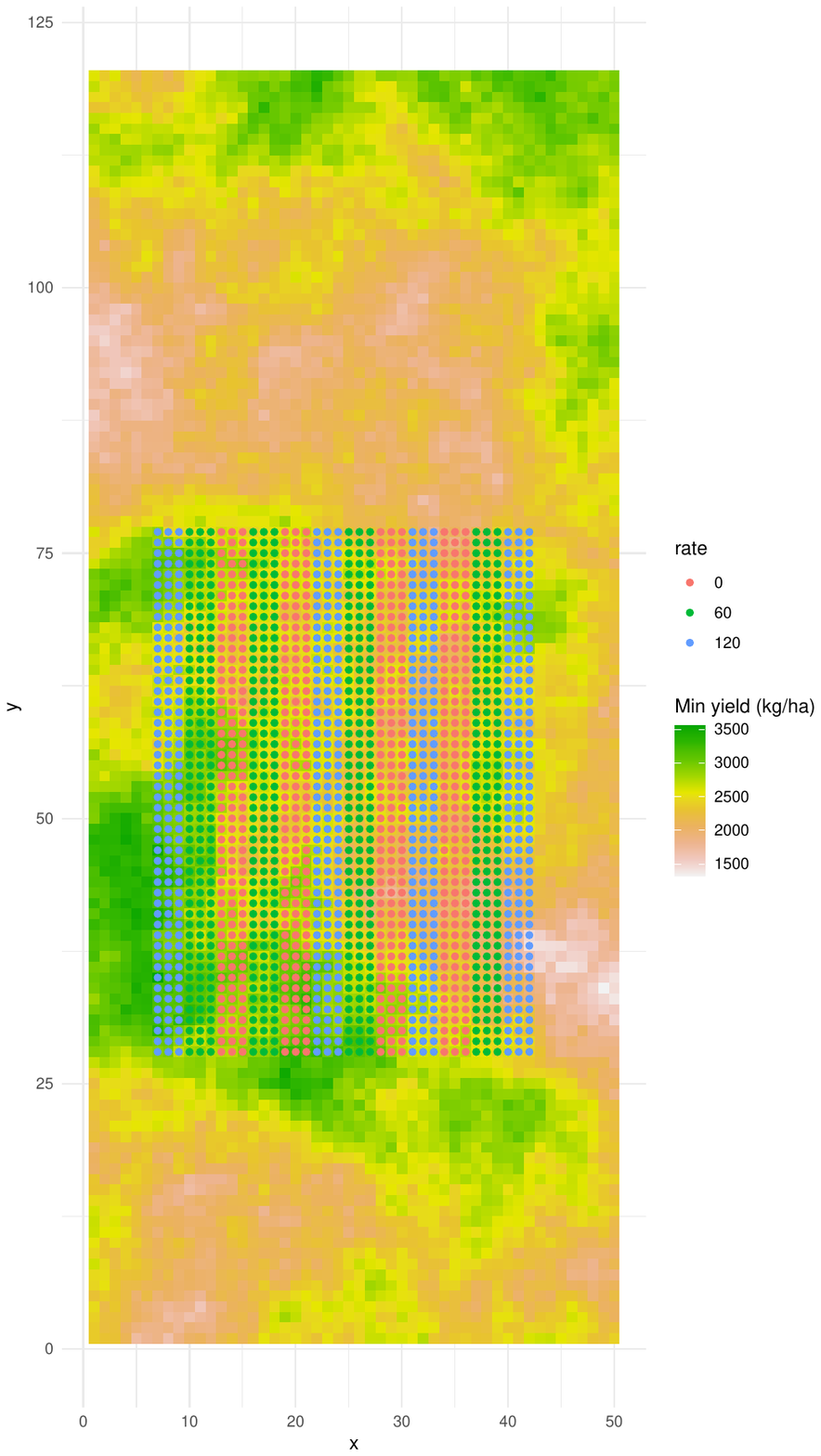}
        \caption{Long-strip layout.}
        \label{fig:farm-strip}
    \end{subfigure}
    \hfill
    \begin{subfigure}[t]{0.48\textwidth}
        \centering
        \includegraphics[width=\linewidth]{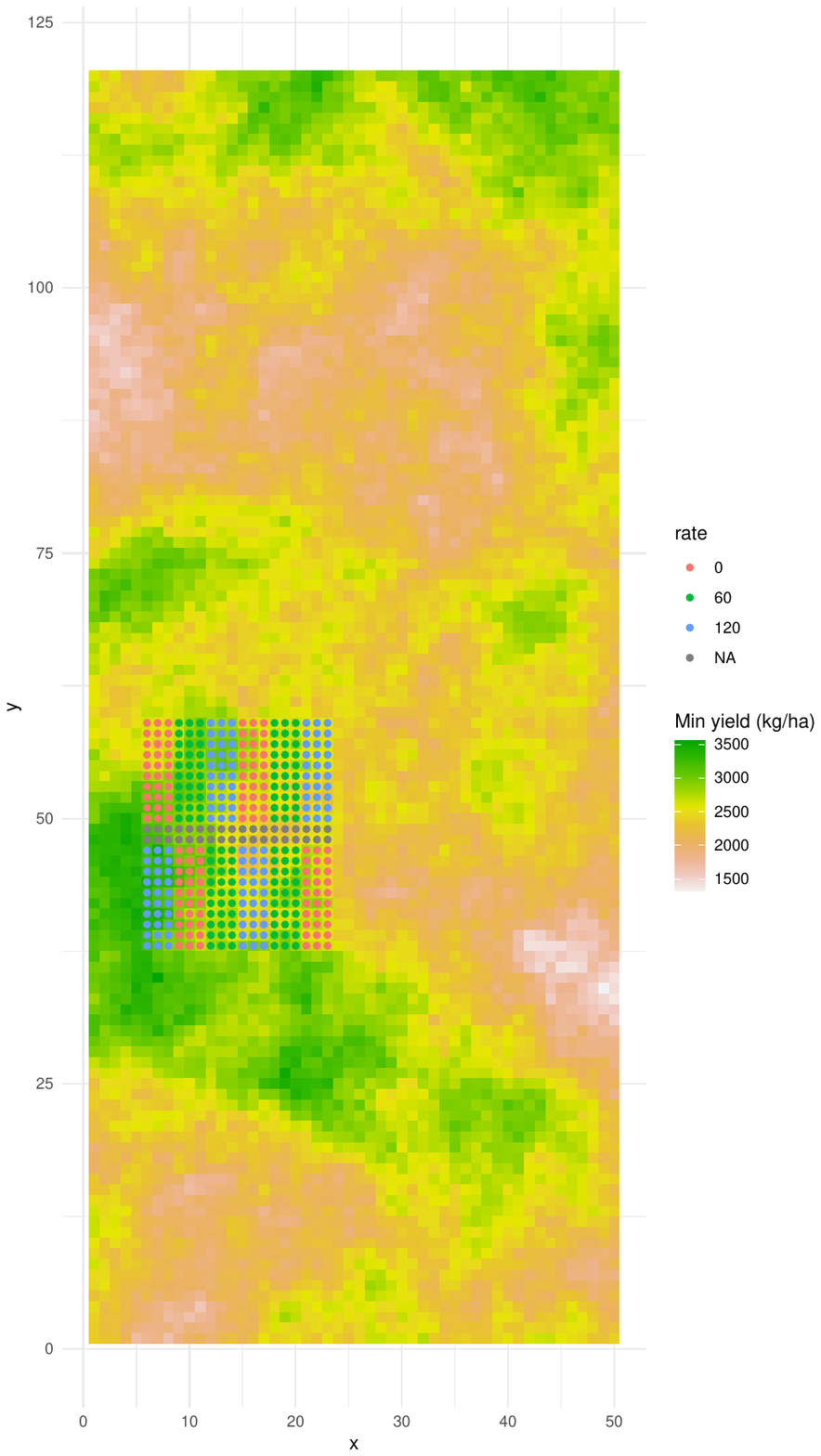}
        \caption{Stacked-replicate layout.}
        \label{fig:farm-stack}
    \end{subfigure}
    \caption{The two trial layouts evaluated in this study, placed on a simulated field under randomised treatment allocation. The background shows the simulated baseline (zero-nitrogen) yield, labelled ``Min yield''; axes are in pixels. Point colours denote the three nitrogen rates. In the long-strip layout (left), each treatment occupies strips spanning the full paddock length. In the stacked-replicate layout (right), treatment sets are placed in the upper and lower halves of the paddock, allowing more replicates to be accommodated within a shorter paddock.}
    \label{fig:farm-layouts}
\end{figure}

\begin{figure}[H]
    \centering
    \includegraphics[width=0.85\textwidth]{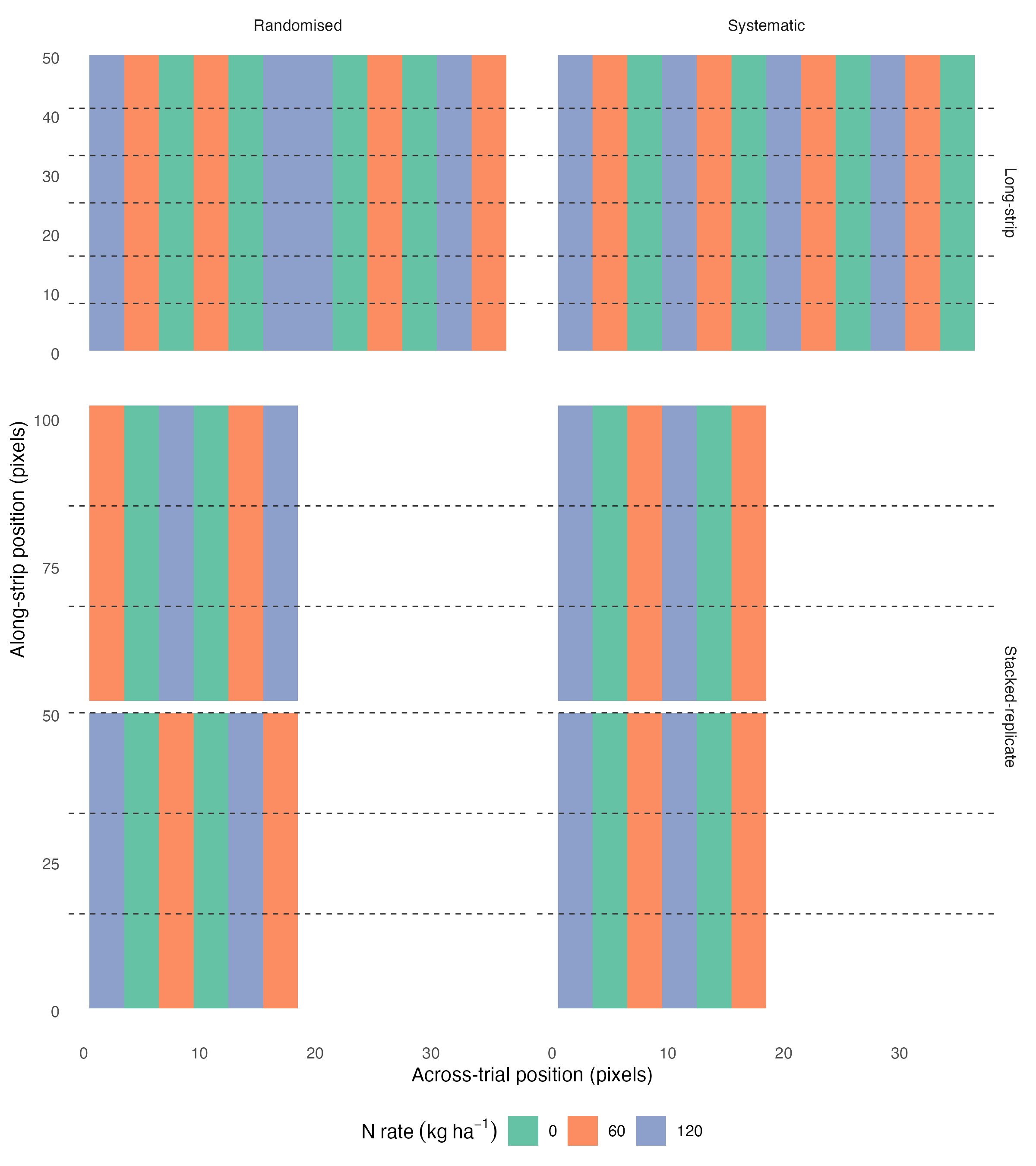}
    \caption{Simulated trial layouts: long-strip (top) and stacked-replicate (bottom) designs under randomised (left) and systematic (right) treatment allocation. Each colour represents a fertiliser rate ($N = 0$, 60, 120\,kg\,ha$^{-1}$). The trial footprint is partitioned into six horizontal PEs (dashed lines) as used in Scenario~B.}
    \label{fig:layout}
\end{figure}

Two layouts were evaluated, both motivated by commercial practice (Figure~\ref{fig:farm-layouts}). In the \textit{long-strip} layout, each treatment is applied to strips running the full length of the paddock, and treatments are replicated across the paddock width. In the \textit{stacked-replicate} layout, two sets of treatment strips are placed end to end in the upper and lower halves of the paddock, separated by a 2-pixel gap, which doubles the number of replicates that fit into a given paddock width. Strips were $w = 3$\,pixels wide in both layouts. Five trial lengths ($L \in \{10, 20, 50, 80, 110\}$\,pixels; for stacked layouts, the length of each block) were crossed with six replication levels ($r \in \{3,4,6,8,10,12\}$ for strip; $r \in \{2,4,6,8,10,12\}$ for stacked) under both randomised RCBD and systematic allocation (Figure~\ref{fig:layout}). In the systematic design the treatment order of the first replicate was repeated in every replicate.

\subsection{Simulation runs}
\label{sec:runs}
The main simulation crossed layout, scenario, PE count, trial length and replication, with $M = 100$ Monte Carlo iterations per cell under the alternative; each iteration was analysed under both randomised and systematic allocation. The Scenario~A grid was repeated in full with $\beta = 0$ to estimate Type I error, and a Scenario~B null simulation ($\beta_k = 0$ in every PE; $M = 100$) was run at a representative design point ($L = 50$ for strips, $L = 20$ for stacked trials, $r = 6$) across all PE counts. The main simulation used the pixels along the centre column of each three-pixel strip. Follow-up simulations, listed in Table~\ref{tab:runs}, examined specific questions on the same field and with the same trial functions. Three of them used all three pixel columns of every strip and introduced the data situations in which aggregation and pixel-level analysis might differ:
\begin{itemize}
    \item \emph{Complete data}: every pixel observed, uniform response.
    \item \emph{Unequal strip lengths}: an along-strip trend whose slope changes across the trial, $t(s) = 250u + 250uv$, where $u$ and $v$ are the trial coordinates scaled to $[-1, 1]$ along and across the strips, and a diagonal paddock edge, so that column $j$ of the footprint is kept to its first $T - 0.5T(j-1)/(W-1)$ rows ($T$ and $W$ are the footprint length and width).
    \item \emph{Gaps within strips}: the same trend, and each plot loses one contiguous block of 10--40\% of its length at a random position (all three columns), as happens when headland turns, blockages or outlier screening remove yield-monitor records \parencite{Sudduth2007Yield, Lyle2014Postprocessing, Vega2019Protocol}.
    \item \emph{Spatially varying response}: the response coefficient is a Gaussian random field, $\beta(s)$ with mean 6, standard deviation 1.8 and exponential range 40\,pixels, drawn anew in every iteration; the target is the response map.
\end{itemize}
Within each design cell these settings share the trial position and allocation, and differ only in trend, missingness or response. In total, more than 3 million models were fitted.

\begin{table}[H]
\centering
\caption{Simulation runs. All runs use the same simulated field (one realisation) unless stated, conditional Wald $F$-tests, and both randomised and systematic allocation unless stated. Strip: long-strip layout; stack: stacked-replicate layout.}
\label{tab:runs}
\small
\begin{tabular}{p{4.2cm} p{6.4cm} p{4.6cm}}
\toprule
Run & Design cells & Purpose \\
\midrule
Main grid & Both layouts, all $L \times r$; Scenario~A ($\beta = 0, 6$) and B ($K = 2$--10); $M = 100$; centre pixel column & Validity, power check, accuracy, PE models (Sections~\ref{sec:calibrate}, \ref{sec:detect} and~\ref{sec:pe}) \\
Length--replication follow-up & Strip, Scenario~A, all $L \times r$; $M = 500$ & Matched-area comparisons (Section~\ref{sec:length-reps}) \\
Pixel models with a strip effect & Scenario~A grid refitted with a random plot effect; same seeds as the main grid & Mechanism of pixel-model validity (Section~\ref{sec:calibrate}) \\
Stacked mechanism & Stack, $L \in \{20, 50\}$, $r \in \{4, 6, 8\}$; null $M = 500$ & Remedies for liberal pixel models \\
Heterogeneity and effect size & Strip $L = 50$, $r = 4$; stack $L = 20$, $r = 6$; 3 realisations $\times$ 5 fields; $\beta = 1.5, 3$ & Robustness, power check (Sections~\ref{sec:detect} and~\ref{sec:robust}) \\
Scenario~B whole-plot & Representative points, $K = 2$--10; null $M = 300$ & Strip effect, multiplicity (Section~\ref{sec:pe}) \\
Pixel diagnostic & Strip $L = 50$, $r = 4$, randomised; $M = 300$; three columns & Covariance family, all four data situations (Section~\ref{sec:pixel}) \\
Elasticity across fields & Both layouts, all $L \times r$; Scenario~A null; $M = 200$; three columns; 3 realisations $\times$ 4 fields & Robustness of the design levers (Section~\ref{sec:robust}) \\
Pixel grid & Both layouts, all $L \times r$; $M = 100$; three columns; complete, unequal lengths, gaps, varying response (strip only) & Design comparisons; when pixel analyses pay off (Sections~\ref{sec:detect}--\ref{sec:layout} and~\ref{sec:pixel}) \\
PE pixel & Strip $L \in \{50, 110\}$, stack $L \in \{20, 50\}$, $r \in \{4, 8\}$, $K \in \{2, 4, 6\}$; $M = 100$; three columns; complete and gaps & Spatial PE models (Section~\ref{sec:pe}) \\
\bottomrule
\end{tabular}
\end{table}

\subsection{Analysis models}
\label{sec:models}
A strip trial differs from a small-plot trial in two ways that shape the analysis. First, yield varies smoothly across the paddock, so neighbouring observations are correlated. Second, the paddock may contain areas that respond differently to the treatment. Linear mixed models (LMM) handle both: random effects and correlated residuals describe the spatial structure, here mainly through separable autoregressive (AR1) residuals along and across strips \parencite{Gilmour1997Accounting}, and fixed effects describe the treatment response. We treat a large strip trial in the same way as a multi-environment trial, in which the field is partitioned into PEs that play the role of environments \parencite{stefanova2023}. Suppose the field is partitioned into $K$ PEs, and the $k$-th PE consists of $n_k$ grid points with response vector ${y}_k$. The LMM for ${Y} = \lbrack{y}^\top_{1}, \ldots, {y}^\top_{K}\rbrack^{\top}$ is
\begin{equation}\label{eq:modelmatrix}
    {Y} = {X}{\tau} + {Z}{u} + {e},
\end{equation}
where ${\tau}$ and ${u}$ are vectors of fixed and random effects with design matrices ${X}$ and ${Z}$, and ${e}$ is the residual vector. The analysis models differ in three places within model~\eqref{eq:modelmatrix}: in what ${Y}$ contains (strip-plot means or individual pixels), in how ${X}\tau$ represents the treatment (a single paddock-wide effect or a separate effect for each PE), and in the covariance assumed for ${u}$ and ${e}$.

``Pixel'' data refers to the gridded yield pixels of each strip, all sharing a common treatment assignment; ``aggregated'' data replaces them with a single strip-plot mean (or one mean per strip and PE). In a strip trial the experimental unit is the strip, because the treatment is applied to the whole strip at once, and the pixels within a strip are repeated measurements of that unit. Treating them as independent replicates is pseudo-replication \parencite{Hurlbert1984Pseudoreplication}: it makes the trial appear to contain far more information than it does, and so produces standard errors that are too small and tests that reject too often. The model suite examines three remedies: aggregating pixels to one value per unit, modelling their correlation with a spatial residual, and including the strip as a random plot effect, which gives the subsampling structure of an RCBD with several measurements per plot. Table~\ref{tab:models-overview} gives a plain-language overview, and Tables~\ref{tab:models-scena} and~\ref{tab:models-scenb} (Appendix~\ref{app:models}) the Scenario~A and Scenario~B models. In the follow-up simulations a suffix \texttt{3} marks models fitted to all three pixel columns (Ag3 and PEag3: means over all three columns; Px3 and PEpx3: all pixels, with \texttt{ar1(column):ar1(row)} over pixel columns 1\,pixel apart). Response maps were estimated with geographically weighted regression (GWR) of yield on nitrogen rate, with a fixed Gaussian kernel of bandwidth 3 or 6\,pixels \parencite{Rakshit2020Novel, evans2020Assessment}, and with the PE models. Analysis used \asreml{} \parencite{butler2023asreml} in \R{} \parencite{RCoreTeam2025R} and the \texttt{GWmodel} package \parencite{Gollini2015GWmodel}.

\begin{table}[H]
\centering
\caption{Plain-language overview of the analysis models. Codes are read from the model's features: \texttt{Ag}/\texttt{Px} denote aggregated strip-plot means / gridded pixels; the \texttt{PE} prefix marks the PE-aware (Scenario~B) models; the residual token is \texttt{0} (none), \texttt{AR1}, \texttt{AR1w} (within-strip), \texttt{dsum} (per-PE) or \texttt{fAR1} (field-wide); \texttt{.plot} adds a random strip (plot) effect; a trailing \texttt{.d}/\texttt{.c} gives the PE random effect (\texttt{diag}/\texttt{corh}); \texttt{.i} marks the interaction (\TPE) parameterisation; and \texttt{3} marks models fitted to all three pixel columns. An asterisk stands for both the \texttt{.d} and \texttt{.c} variants. Validity refers to Type I error under randomisation (Sections~\ref{sec:calibrate}, \ref{sec:pixel} and~\ref{sec:pe}).}
\label{tab:models-overview}
\small
\begin{tabular}{p{4.4cm} p{11.2cm}}
\toprule
Model(s) & Role / note \\
\midrule
Ag.0, Ag3.0 & Strip-plot means, no spatial term (classical RCBD on plot means): valid under randomisation; liberal under systematic allocation \\
Ag.AR1, Ag3.AR1 & Strip-plot means $+$ AR1 spatial residual: most precise with complete data; liberal when strips contain gaps \\
Px.0 & Gridded pixels, no spatial term: invalid (pseudo-replication) \\
Px.AR1, Px.AR1w & Gridded pixels $+$ two-dimensional / within-strip AR1 residual, no plot term: conservative on long strips; liberal on randomised stacked layouts \\
Px.AR1.plot, Px3.AR1.plot & Gridded pixels $+$ AR1 residual $+$ random strip (plot) effect: valid under randomisation, liberal under systematic allocation; most precise when strips contain gaps \\
PEag.0.* & PE-aware, aggregated, no spatial residual \\
PEpx.0.* & PE-aware, gridded pixels, no spatial residual: invalid (pseudo-replication) \\
PEag.dsum.*, PEpx.dsum.* & PE-aware $+$ one-dimensional per-PE (dsum) spatial residual (aggregated / pixel) \\
PEpx.dsum2D.*, PEpx.dsumAd.* & PE-aware, pixels $+$ two-dimensional per-PE residual; the adaptive form chooses each PE's structure from its own geometry \\
PEag.fAR1.*, PEpx.fAR1.* & PE-aware $+$ field-wide AR1 residual (aggregated / pixel) \\
GWR & Geographically weighted regression of pixel yield on rate; local response maps \\
\bottomrule
\end{tabular}
\end{table}

\subsection{Tests and performance measures}
\label{sec:measures}

\paragraph{Tests.} All fixed-effect tests were conditional Wald $F$-tests, obtained in \asreml{} with \texttt{wald(fit,} \texttt{denDF = "default",} \texttt{ssType = "conditional")}. An $F$-test with estimated denominator degrees of freedom accounts for the small number of experimental units in an aggregated analysis, whereas a Wald chi-square test treats the variance parameters as known and is therefore liberal when there are few units. A conditional test assesses each term after adjusting for every other term that does not contain it; this matters for the PE-aware models, in which each PE has its own rate term (\texttt{at(zone):rate}). To isolate the effect of the test itself, the full main grid was also run with \asreml{}'s default incremental Wald chi-square tests (the \emph{incremental run}); because the test affects only $p$-values, this run also confirmed convergence rates and estimates within Monte Carlo error (Section~\ref{sec:test-form}).

\paragraph{Validity (Type I error).} Type I error is the proportion of null iterations in which the test of the treatment term rejected at $\alpha = 0.05$. With $M = 100$ iterations, the Monte Carlo standard error (SE) of a single-cell rate near 5\% is about 2.2 percentage points, so classification is based on rates pooled over cells. A model is \emph{liberal} when its pooled rate exceeds 5\% by more than two SE and \emph{conservative} when it falls below 5\% by more than two SE. A liberal model reports a difference whether or not one exists, so its power carries no information; Type I error is therefore a precondition, and designs are compared only through analyses that pass it. For PE-level tests the family-wise error rate (FWER) is the proportion of null trials in which at least one PE test rejected.

\paragraph{Minimum detectable response.} At $\beta = 6$, almost every valid analysis of almost every design reached $\geq 99\%$ power (Appendix~\ref{app:figures}), so power cannot rank designs. Designs were therefore scored by the minimum detectable response (MDR): the smallest response coefficient $\beta$ (kg grain per kg N) that the rate test detects with 80\% power at $\alpha = 0.05$. For each design cell (layout $\times$ allocation $\times$ model $\times$ $L$ $\times$ $r$), let $\hat{\bm\Sigma}$ be the empirical covariance, over iterations, of the estimation errors of $(\hat\Delta_{60}, \hat\Delta_{120})$. With $\bm v = (60, 120)^\top$, the noncentrality of the 2-df rate test at response $\beta$ is $\beta^2 q$, where $q = \bm v^\top \hat{\bm\Sigma}^{-1} \bm v$, and
\begin{equation}\label{eq:mdr}
\mathrm{MDR} = \sqrt{\lambda_{80}(\nu)/q}, \qquad \Pr\{F_{2,\nu}(\lambda_{80}) > F_{0.95;\,2,\nu}\} = 0.8,
\end{equation}
with denominator degrees of freedom $\nu = 2(r-1)$, the RCBD error degrees of freedom. MDR is reported as the 0--120\,kg\,N yield gain it implies, $120\,\mathrm{MDR}$, expressed as a percentage of mean yield (the \emph{detectable gain}). For PE-level tests the MDR was computed for each PE, with $\nu = 2(r_k - 1)$ where $r_k$ is the number of replicates present in PE $k$, and averaged over PEs. 

Because the treatment effect is additive and REML variance parameters are invariant to the fixed effects, the distribution of the estimation errors does not depend on $\beta$; null and alternative runs were therefore pooled. In the main grid they were pooled with the independent $M = 500$ run of the strip Scenario~A grid, giving 190--700 fits per cell; in the three-column pixel grid, which is used for the design comparisons, each cell had about 200 fits. Two checks support the pooling: the ratio of error SDs between null and alternative runs had median 1.00 (90\% of cells 0.84--1.18), and between the $M = 500$ run and the main run median 1.02 (interquartile range 0.97--1.07). Confidence intervals for MDR and for ratios of MDRs were obtained from bootstrap resamples of iterations within each cell (1{,}000 in the main grid, 500 in the follow-up runs). The MDR measures the precision of a design and analysis; whether the test of that analysis realises it depends on the calibration of the test, which was checked against simulated power at $\beta = 1.5$, 3 and 6 (Section~\ref{sec:detect}).

\paragraph{Estimation and mapping accuracy.} Economic decisions depend on the size of the effect, not only on whether it is significant. Accuracy was measured by the relative absolute deviation (RAD) of each estimated contrast from its true value, $\mathrm{RAD} = |\hat{\Delta} - \Delta| / \Delta$, reported as a median over iterations. For the spatially varying response, each method's local response estimate was compared with $\beta(s)$ at every pixel, and the root mean squared error (map RMSE) was averaged over iterations. For Scenario~B, rate detection power is the proportion of PE-level rate tests that rejected, averaged over PEs, and \TPE power refers to the test of the PE-by-rate interaction. All measures are computed over iterations in which the model converged.

\section{Results}
\label{sec:results}

The results follow the study logic: which analyses are valid (Section~\ref{sec:calibrate}); what a design can detect (Section~\ref{sec:detect}); the four design elements in turn (Sections~\ref{sec:length-reps}--\ref{sec:layout}); robustness to the field (Section~\ref{sec:robust}); when pixel-level analyses pay off (Section~\ref{sec:pixel}); PE-varying responses (Section~\ref{sec:pe}); and a design table that brings them together (Section~\ref{sec:framework}).

\subsection{Calibrating the instrument: which analyses give valid tests}
\label{sec:calibrate}

The non-spatial pixel model Px.0 rejected a true null hypothesis in 38\% of randomised long-strip trials, and in 32--40\% across the other layout and allocation combinations (Table~\ref{tab:typeI}; Figure~\ref{fig:typeI}, Appendix~\ref{app:figures}). This is pseudo-replication: all pixels within a strip share the same treatment, so treating them as independent observations overstates the effective sample size. A trial analysed this way would report a treatment difference in more than one in three cases where none exists. Adding the strip as a random plot effect to the same pixel model removed the problem: in balanced data the resulting subsampling model reproduced the Ag.0 analysis of strip means (Type I error 4.2\% against 4.1\% on randomised long strips, and 13.2\% against 12.8\% on systematic stacked layouts). It is the term for the experimental unit, not aggregation as such, that restores validity.

On long strips, every analysis other than Px.0 controlled Type I error. Ag.0 rejected in 4.1\% of randomised trials and Ag.AR1 in 2.4\%. The pixel-level spatial models fitted to the centre column were strongly conservative (0.3--0.5\%), with or without a plot term. Under systematic allocation Ag.0 rose to 5.9\% ($\pm 0.4$ SE), with individual cells up to 16\%; the spatial models of the main grid were unaffected.

Stacked layouts were less forgiving, in three ways. Under systematic allocation, Ag.0 rejected in 12.6\% of null trials: the same treatment order is repeated in the upper and lower blocks, so smooth positional variation in yield masquerades as a treatment effect when the model ignores space. The aggregated spatial model Ag.AR1 was liberal at low replication (22.0\% at $r = 2$; 7.7--8.1\% at $r = 4$--6) and reached the nominal level from $r = 8$ ($\leq 4.7\%$). Under randomised allocation the pixel-level spatial models Px.AR1 and Px.AR1w rejected in 13\% of null trials (6--22\% across cells). The cause of this last result is the missing experimental-unit term. In the stacked pixel models the strip is represented only through the AR1 residual. Adding a random plot effect (Px.AR1.plot) reduced Type I error to 6.1\% over the full grid (2--13\% across cells) and to 5.9\% in the targeted run with $M = 500$, and fitting the residual separately within each block reduced it to 1.3\%.

\begin{table}[H]
\centering
\caption{Type I error rates (\%) for the Scenario A models under the null hypothesis (true treatment effect $= 0$), by layout and allocation, main grid (centre pixel column). Each entry is the mean over all trial length $\times$ replication cells ($M = 100$ per cell; Monte Carlo SE of each mean $\leq 0.9$ percentage points), with the range across cells in brackets. Nominal rate is 5\%.}
\label{tab:typeI}
\begin{tabular}{lcccc}
\toprule
 & \multicolumn{2}{c}{Long-strip} & \multicolumn{2}{c}{Stacked-replicate} \\
\cmidrule(lr){2-3}\cmidrule(lr){4-5}
Model & Randomised & Systematic & Randomised & Systematic \\
\midrule
Ag.0    & 4.1 (1--9)   & 5.9 (0--16)  & 5.1 (0--15)  & 12.6 (3--20) \\
Ag.AR1  & 2.4 (0--6)   & 3.0 (0--9)   & 8.0 (0--27)  & 4.6 (0--12)  \\
Px.0    & 38.1 (21--57) & 31.6 (19--45) & 37.8 (21--54) & 40.2 (15--62) \\
Px.AR1  & 0.5 (0--3)   & 0.4 (0--3)   & 13.0 (6--22) & 0.9 (0--4)   \\
Px.AR1w & 0.4 (0--3)   & 0.3 (0--2)   & 13.0 (5--22) & 0.3 (0--3)   \\
Px.AR1.plot & 0.3 (0--2) & 0.4 (0--3) & 6.1 (2--13)  & 0.5 (0--4)   \\
\bottomrule
\end{tabular}
\end{table}

The form of the test also matters for the aggregated analyses, which have few degrees of freedom. On six null design cells, the Type I error of Ag.0 was 8.7\% under a Wald chi-square test and 5.2\% under the $F$-test, whereas that of Px.0 was unchanged (36.7\% and 36.1\%), because with thousands of pixels the two reference distributions coincide. With all three pixel columns the same pattern held (complete data, randomised allocation): Ag3.0 rejected in 5.0\% of null trials on long strips and 5.2\% on stacked layouts, Ag3.AR1 in 1.5\% and 6.8\% (valid on stacked layouts from $r = 8$), and the pixel model with a strip effect, Px3.AR1.plot, in 4.5\% and 3.7\%. The designs in Sections~\ref{sec:detect}--\ref{sec:layout} are judged with these three analyses, fitted to all three pixel columns, with conditional $F$-tests.

\subsection{What a strip trial can detect}
\label{sec:detect}

\begin{figure}[H]
    \centering
    \includegraphics[width=\textwidth]{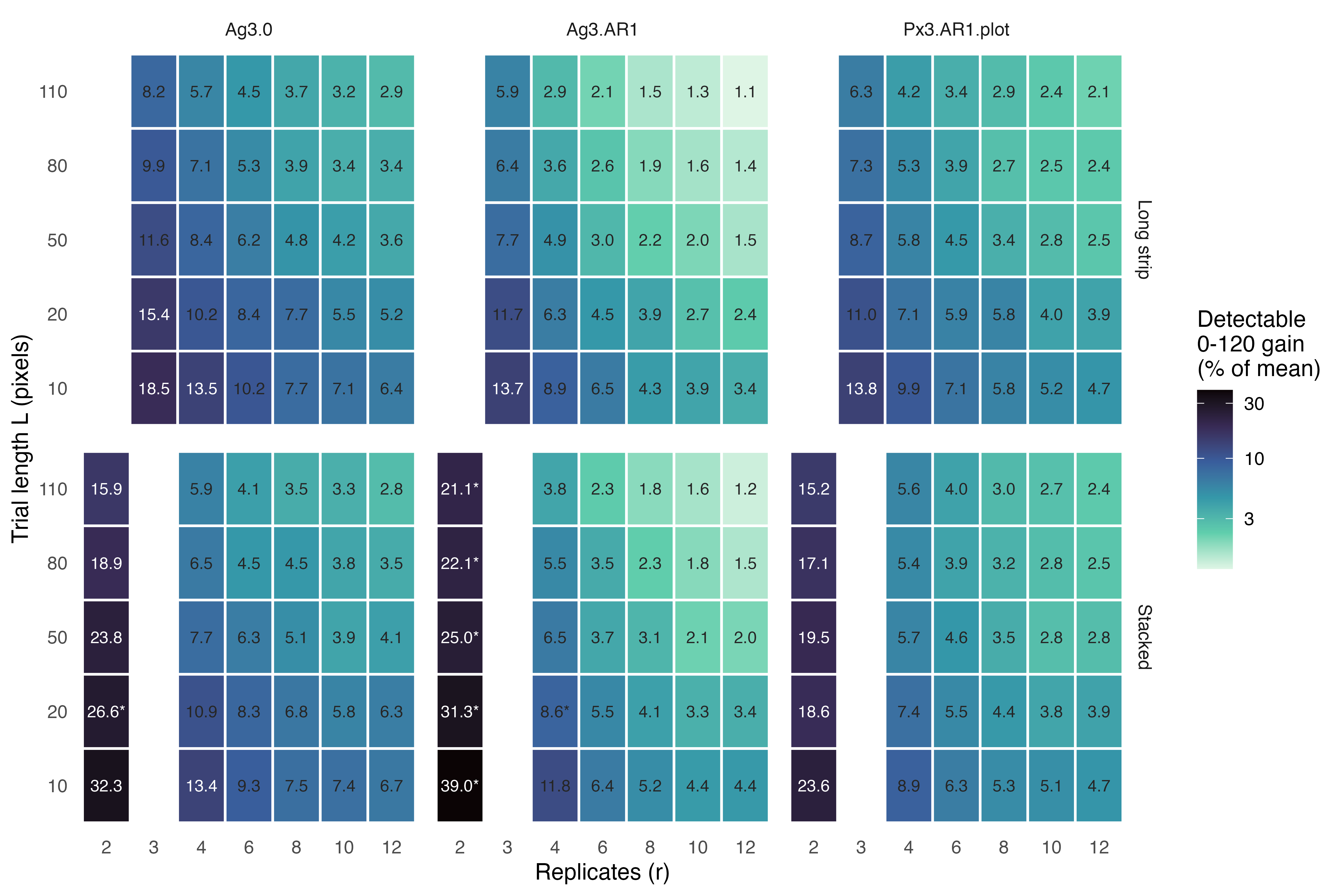}
    \caption{Replication lowers the detectable gain faster than trial length, and an AR1 residual on strip means lowers it further. Detectable gain: the 0--120\,kg\,N yield gain (as \% of mean yield) that the rate test detects with 80\% power at $\alpha = 0.05$, by trial length and replication, for three valid analyses using all three pixel columns, under randomised allocation (Equation~\ref{eq:mdr}; complete data; about 200 fits per cell, null and alternative pooled). Lower is better. An asterisk marks a cell in which the test of that analysis was liberal under the null in that cell alone (rate more than two SE above 5\%, about 100 null fits); with so few fits per cell, a moderate excess such as that of Ag3.AR1 on stacked layouts at $r = 4$--6 (pooled 6.0--8.9\%; Section~\ref{sec:layout}) is flagged in only some cells. The simulated effect corresponds to 28.8\%.}
    \label{fig:mdr}
\end{figure}

At the simulated response (a 0--120\,kg\,N gain of 720\,kg\,ha$^{-1}$, 29\% of mean yield), the aggregated analyses reached $\geq 99\%$ power from $r = 3$ on long strips and from $r = 4$ on stacked layouts (Appendix~\ref{app:figures}), so power cannot distinguish the designs. The detectable gain does (Figure~\ref{fig:mdr}; three pixel columns, complete data). For Ag3.AR1 on randomised long strips it ranged from 13.7\% of mean yield ($L = 10$, $r = 3$) to 1.1\% ($L = 110$, $r = 12$), for Ag3.0 from 18.5\% to 2.9\%, and for Px3.AR1.plot from 13.8\% to 2.1\%. At $L = 50$ and $r = 4$, a common commercial configuration, a trial analysed with Ag3.AR1 detects a gain of 4.9\% of mean yield (about 120\,kg\,ha$^{-1}$) with 80\% power, about one sixth of the simulated effect; this is why power saturated. On stacked layouts with $r \geq 4$ the detectable gain of Ag3.AR1 ranged from 1.2 to 11.8\%; at $r = 2$ no valid analysis detected less than 15\%. Averaging the centre column only, as in the main grid, gave a larger detectable gain for the aggregated spatial analysis: the ratio of Ag.AR1 to Ag3.AR1 had a median of 1.22 on long strips and 1.16 on stacked layouts (Section~\ref{sec:pixel}).

Whether an analysis realises this precision depends on the calibration of its test. This was checked in the main grid (centre column) at two smaller response coefficients (Table~\ref{tab:power-check}, Appendix~\ref{app:figures}). For the aggregated analyses under randomisation, power predicted from the MDR calculation agreed with simulated power to within 4 percentage points (Ag.AR1 at $\beta = 1.5$: 94.0\% predicted against 92.7\% observed on long strips, 91.1\% against 91.6\% on stacked layouts; Ag.0 on stacked layouts: 65.3\% against 64.7\%). For these analyses the MDR can be read directly as what a trial detects. Two departures are informative. Under systematic allocation, Ag.0 detected more often than predicted (70.3\% against 55.1\% on long strips; 74.0\% against 44.5\% on stacked layouts), which is the counterpart of its inflated Type I error. The centre-column pixel model fell far short of its prediction on long strips (Px.AR1.plot: 18.4\% observed against 73.0\% predicted at $\beta = 1.5$; 73.7\% against 99.9\% at $\beta = 3$): its estimates were at least as precise as those of Ag.0, but its standard errors were too large, so its conservative test discarded most of that precision. Fitted to all three columns, the same model had a calibrated test on long strips (Type I error 4.5\%; Section~\ref{sec:calibrate}).

The analysis is itself a design lever of the same size as replication. Ag3.AR1 had a smaller MDR than Ag3.0 in every long-strip cell, by a median factor of 1.98 (range 1.32--2.63), and on stacked layouts by a median factor of 1.61 (0.75--2.36); that is, the autoregressive residual on strip means lowered the detectable response by a median 49\% on long strips and 38\% on stacked layouts. At $L = 50$ and $r = 4$ on long strips, Ag3.AR1 detects 4.9\% where Ag3.0 detects 8.4\%; Ag3.0 needs eight replicates (4.8\%) to match it.

\subsection{Length and replication}
\label{sec:length-reps}

Both longer strips and more replicates reduced the detectable gain, but at very different rates (Table~\ref{tab:elasticity}). Over the replication levels shared by both layouts ($r = 4$--12), a log-linear fit of MDR on $\log L$ and $\log r$ explained 97--99\% of the variation across cells. If every pixel were an independent observation, the error SD would fall as $L^{-0.5}$ and $r^{-0.5}$. Length fell short of this: MDR scaled as $L^{-0.24}$ to $L^{-0.45}$, because pixels along a strip are correlated and each additional pixel adds less new information than the last. In the terms of the law of \textcite{Smith1938Empirical}, in which the variance of a plot mean falls as plot size to the power $-b$, the length elasticity of the error SD for plain strip means (Ag3.0, $-0.34$) corresponds to $b \approx 0.7$. Replication approached or exceeded the independent-observation rate: the error SD scaled as $r^{-0.41}$ to $r^{-0.84}$, and MDR, which also gains error degrees of freedom with each replicate, as $r^{-0.60}$ to $r^{-1.02}$. These values refer to the base field; Section~\ref{sec:robust} repeats the estimation on four heterogeneity structures.

\begin{table}[H]
\centering
\caption{Elasticity of precision with respect to trial length and replication (three pixel columns, complete data, randomised allocation; $r = 4$--12; all five lengths). Entries are the coefficients of $\log L$ and $\log r$ in a linear model for the log error SD of $\hat\Delta_{120}$ or the log MDR; more negative is a stronger lever. Under independent pixels both SD elasticities would be $-0.5$. The last two columns give, over the matched-area pairs of each layout (both allocations), the number of pairs in which the design with more, shorter replicates had the smaller MDR, and the median ratio of the two MDRs (fewer-longer / more-shorter).}
\label{tab:elasticity}
\small
\begin{tabular}{llcccccc}
\toprule
 & & \multicolumn{2}{c}{Error SD} & \multicolumn{2}{c}{MDR} & \multicolumn{2}{c}{Matched area} \\
\cmidrule(lr){3-4}\cmidrule(lr){5-6}\cmidrule(lr){7-8}
Layout & Model & $L$ & $r$ & $L$ & $r$ & Reps win & Median ratio \\
\midrule
Long strip & Ag3.0        & $-0.34$ & $-0.51$ & $-0.33$ & $-0.69$ & 10/10 & 1.53 \\
           & Ag3.AR1      & $-0.45$ & $-0.74$ & $-0.45$ & $-0.92$ & 10/10 & 1.81 \\
           & Px3.AR1.plot & $-0.33$ & $-0.50$ & $-0.33$ & $-0.68$ & 10/10 & 1.56 \\
Stacked    & Ag3.0        & $-0.34$ & $-0.41$ & $-0.34$ & $-0.60$ & 12/12 & 1.84 \\
           & Ag3.AR1      & $-0.44$ & $-0.84$ & $-0.44$ & $-1.02$ & 12/12 & 2.67 \\
           & Px3.AR1.plot & $-0.24$ & $-0.49$ & $-0.24$ & $-0.68$ & 12/12 & 1.79 \\
\bottomrule
\end{tabular}
\end{table}

At a fixed plot area this has a direct consequence. Among design cells with the same product $L \times r$ (the same plot area per treatment), the design with more, shorter replicates had the smaller MDR in all 66 matched comparisons (three valid analyses, both allocations, 11 pairs across the two layouts), and the bootstrap intervals did not overlap in 61 of them. On randomised long strips with Ag3.AR1, for example, $L = 20$ with $r = 12$ detects 2.4\% where $L = 80$ with $r = 3$ detects 6.4\%, and $L = 20$ with $r = 10$ detects 2.7\% where $L = 50$ with $r = 4$ detects 4.9\%. The median MDR ratio was 1.53--1.81 on long strips and larger on stacked layouts, where the matched pairs include $r = 2$.

Estimation accuracy told the same story. In the $M = 500$ strip run (centre column), allocating a fixed $L \times r$ to more replicates gave lower RAD than allocating it to greater length in 74--84\% of matched comparisons (Ag.AR1, randomised: 79\% for $\hat{\Delta}_{60}$, 74\% for $\hat{\Delta}_{120}$; Ag.AR1, systematic: 84\%; Px.AR1, randomised: 79\%); at $L \times r \approx 240$, for instance, $L = 20$ with $r = 12$ achieved median RAD of 4.8\% against 6.1\% for $L = 80$ with $r = 3$. Length nonetheless kept improving accuracy over the whole range evaluated: for Ag.AR1 under randomisation, median RAD for $\hat{\Delta}_{60}$ fell from 8.6\% at $L = 10$ to 4.1\% at $L = 50$ and 2.8\% at $L = 110$, and at $r = 3$ from 14\% to 5\%.

Both intuitions therefore hold, but they are not equal. More replicates and longer strips both make a trial more informative; replication buys roughly twice as much per unit of plot area. Because $L \times r$ measures area rather than cost, the practical reading depends on what constrains the trial: where strip length is set by the paddock and costs nothing extra, strips should run its full length, and where the choice is between more replicates and longer strips, replicates are the better investment.

\subsection{Randomised or systematic allocation}
\label{sec:allocation}

Randomisation did not make the analyses of strip means more precise. On long strips, systematic allocation had a slightly smaller MDR than randomised allocation for the analyses of means (median ratio systematic/randomised 0.88 for Ag3.0 and 0.94 for Ag3.AR1, with the bootstrap interval favouring systematic allocation in 13 and 8 of 30 cells and randomised allocation in none). On stacked layouts the aggregated spatial analysis was also slightly more precise under systematic allocation (0.90), but the non-spatial analysis was less precise (1.14; randomised better in 13 of 30 cells). The pixel-level model was less precise under systematic allocation on both layouts (1.08 on long strips and 1.35 on stacked layouts; randomised better in 8 and 23 of 30 cells).

What randomisation buys is validity. Under systematic allocation the non-spatial analysis of means was liberal at low replication on long strips (Ag3.0 8.0--8.2\% at $r = 3$--6) and at every replication from $r = 4$ on stacked layouts (11.8--16.2\%; 12.7\% pooled). The pixel-level model was liberal on both layouts (Px3.AR1.plot 12.6\% on long strips and 17.2\% on stacked layouts), whereas the aggregated spatial analysis remained valid or conservative (Ag3.AR1 1.8\% and 3.5\%). The main grid showed the same pattern with the centre column (Ag.0 5.9\% and 12.6\%; Table~\ref{tab:typeI}). On fields with an added linear trend (Section~\ref{sec:robust}), systematic Ag.0 rejected in 10.2\% of null trials on long strips and 12.2\% on stacked layouts, against 4.4\% and 3.8\% under randomisation. The base fields gave similar rates (10.4\% and 14.2\%), so the smooth background heterogeneity of the field is enough to align with a repeated treatment order, without an explicit trend. Averaged over trial positions, the estimated contrast was not biased under either allocation (mean error within 6\,kg\,ha$^{-1}$ against an error SD of 31--77\,kg\,ha$^{-1}$). The problem is that, at a given position, a systematic order aligns with the local yield pattern and the standard error does not account for it.

The third intuition is therefore conditional. For testing a treatment contrast, randomisation is insurance for whichever analysis is eventually applied, and only a spatial residual on strip means remained valid without it. When the aim is a response map rather than a test, systematic allocation is the better choice (Section~\ref{sec:pixel}).

\subsection{Long-strip or stacked layout}
\label{sec:layout}

The comparison depends on what is held fixed. With the same plots (the same $L$ and $r$, so the same number of strips of the same size), the long-strip arrangement was almost never less precise than the stacked one: the stacked layout was significantly more precise in only 2 of 150 comparisons (three analyses, two allocations, 25 cells each). For Ag3.0 and Px3.AR1.plot under randomisation the two arrangements were equivalent (median MDR ratio stacked/long-strip 1.01 and 1.02); for Ag3.AR1 the long strip was better by a median factor of 1.22 (bootstrap interval excluding 1 in 17 of 25 cells); and under systematic allocation the long strip was better by 1.25--1.34 for every analysis.

On the same paddock footprint, however, the stacked layout fits twice the replicates at half the strip length, and because replication is the stronger lever, this can outweigh the arrangement penalty. Over nine pairs of designs with $r_{\text{stacked}} = 2r_{\text{strip}}$ and a footprint within $-16\%$ to $+10\%$ of the long-strip footprint, the stacked layout had the smaller MDR for every analysis under randomisation (median ratio 0.80 for Ag3.0, 0.83 for Ag3.AR1 and 0.77 for Px3.AR1.plot; stacked better in 5--7 of 9 pairs, long strip better in none). For example, a stacked trial of two blocks of $L = 20$ with $r = 8$ detects 4.1\% (Ag3.AR1) on a footprint 16\% smaller than a long-strip trial with $L = 50$ and $r = 4$, which detects 4.9\%.

The stacked layout carries conditions that the long strip does not. Its non-spatial and pixel-level analyses are liberal under systematic allocation (12.7\% and 17.2\%). Its aggregated spatial analysis needs $r \geq 8$ to hold the nominal Type I error under randomisation (Ag3.AR1 20.7\% at $r = 2$, 8.9\% at $r = 4$, 6.0\% at $r = 6$ and $\leq 2.4\%$ from $r = 8$). Any pixel-level analysis also needs a plot term (Section~\ref{sec:calibrate}). The fourth intuition therefore holds plot for plot, and fails paddock for paddock: when a paddock is too short or too narrow for the replication wanted, stacking is a legitimate way to buy replicates, provided the treatments are randomised.

\subsection{Robustness to the field}
\label{sec:robust}

The design results above derive from one realisation of one heterogeneity structure. Table~\ref{tab:robust} repeats the representative design points on three independent realisations of the base field and of four alternative fields. The validity of the instruments held across fields: Ag.0 under randomisation stayed at 3.8--5.3\% in every setting, Px.0 remained invalid (25.6--53.8\%), and systematic allocation raised Ag.0 above the nominal rate in every setting except the short-range field (7.3--16.0\%, against 2.2--5.6\% at range 10). Two qualifications emerged. First, a nugget, which adds independent cell-level noise, roughly halved the inflation of Px.0 (from 40\% to 26\% on long strips). It also made the centre-column pixel-level spatial models liberal under systematic allocation (Px.AR1.plot 6.0--7.1\%; Px.AR1 without a plot term 11.3--16.0\%), so how severe the pixel-level problems are depends on how much of the within-strip variation is spatially structured. Second, Ag.AR1 on stacked layouts at $r = 6$ remained above nominal in most settings (6.5--11.3\%), consistent with the $r \geq 8$ requirement. The precision ranking of the analyses held everywhere: the error SD of Ag.AR1 was 11--40\% smaller than that of Ag.0 in every setting. Across individual realisations, Type I error rates for Ag.0 under randomisation varied from 2.0 to 7.3\%, within the range expected from Monte Carlo error.

\begin{table}[H]
\centering
\caption{Type I error (\%) at the representative design points (long strip $L = 50$, $r = 4$; stacked $L = 20$, $r = 6$) on three independent realisations of each field (about 450 null fits per entry; SE $\approx 1$ percentage point near 5\%). Base: exponential, range parameter 30 pixels. Range 10/60: range parameter 10 or 60 pixels. Nugget: 25\% of the sill as independent noise. Trend: linear trend added to the base field. The last row of each block gives the ratio of the error SD of $\hat\Delta_{120}$ for Ag.AR1 to that for Ag.0 (randomised).}
\label{tab:robust}
\small
\begin{tabular}{llccccc}
\toprule
Layout & Model (allocation) & Base & Range 10 & Range 60 & Nugget & Trend \\
\midrule
Long strip & Ag.0 (randomised)        & 4.0  & 4.9  & 5.3  & 4.0  & 4.4 \\
           & Ag.0 (systematic)        & 10.4 & 2.2  & 11.3 & 7.3  & 10.2 \\
           & Ag.AR1 (randomised)      & 4.5  & 2.9  & 2.9  & 4.7  & 2.9 \\
           & Px.0 (randomised)        & 40.4 & 40.7 & 39.6 & 25.6 & 38.0 \\
           & Px.AR1.plot (randomised) & 0.0  & 0.0  & 0.4  & 2.7  & 0.4 \\
           & Px.AR1.plot (systematic) & 1.1  & 0.2  & 1.8  & 6.0  & 0.2 \\
           & SD ratio Ag.AR1/Ag.0     & 0.60 & 0.72 & 0.60 & 0.80 & 0.64 \\
\midrule
Stacked    & Ag.0 (randomised)        & 4.0  & 4.2  & 3.8  & 3.8  & 3.8 \\
           & Ag.0 (systematic)        & 14.2 & 5.6  & 16.0 & 7.6  & 12.2 \\
           & Ag.AR1 (randomised)      & 7.6  & 8.1  & 6.5  & 11.3 & 8.0 \\
           & Px.0 (randomised)        & 49.1 & 44.9 & 48.4 & 28.0 & 48.2 \\
           & Px.AR1.plot (randomised) & 5.8  & 4.7  & 7.8  & 3.3  & 6.7 \\
           & Px.AR1.plot (systematic) & 1.3  & 0.4  & 1.3  & 7.1  & 0.2 \\
           & SD ratio Ag.AR1/Ag.0     & 0.79 & 0.82 & 0.74 & 0.89 & 0.75 \\
\bottomrule
\end{tabular}
\end{table}

The design levers were then re-estimated over the full $L \times r$ grid on four of these fields (base, range 10, range 60 and nugget; three realisations each, three pixel columns, null fits, $M = 200$ per cell; Table~\ref{tab:elas-fields}). Replication was the stronger lever in every field, for every layout and analysis: the MDR scaled as $r^{-0.63}$ to $r^{-1.13}$ and as $L^{-0.20}$ to $L^{-0.46}$. The length elasticity depended on the correlation range. For plain strip means on long strips it was $-0.40$ in the short-range field and $-0.28$ in the long-range field, which corresponds to a Smith exponent $b$ of 0.79 and 0.55: the longer the range, the more of each additional length of strip repeats information already collected. On the three realisations of the base field, the length elasticities agreed with Table~\ref{tab:elasticity} to within 0.01 and the replication elasticities to within 0.10, which indicates the Monte Carlo precision of the latter.

The design comparisons held in every field. At equal plot area, the design with more, shorter replicates had the smaller MDR in all 66 matched comparisons in each field (bootstrap intervals not overlapping in 60--62; median ratio 1.59--1.77). With the same plots, the long strip was at least as precise as the stacked layout (median MDR ratio stacked/long-strip 0.99--1.06 for Ag3.0 and Px3.AR1.plot and 1.09--1.23 for Ag3.AR1 under randomisation; 1.12--1.31 under systematic allocation). On the same footprint, the stacked layout had the smaller MDR under randomisation for every analysis in every field (median ratio 0.70--0.89; long strip better in none of the nine pairs). The AR1 residual on strip means reduced the MDR by a median factor of 1.57--1.99 on long strips and 1.39--1.67 on stacked layouts. The validity conditions were also stable. Ag3.AR1 on randomised stacked layouts was liberal at $r = 2$ in every field (20.5--21.5\%), and at $r = 8$ its rate was 3.0--5.8\%. Under systematic allocation, the inflation of Ag3.0 and Px3.AR1.plot grew with the correlation range: on stacked layouts they rejected in 5.2\% and 8.6\% of null trials in the short-range field, and in 20.0\% and 24.8\% in the long-range field.

\begin{table}[H]
\centering
\caption{Elasticity of the MDR with respect to trial length ($L$) and replication ($r$) on four heterogeneity structures (three realisations each; three pixel columns, complete data, randomised allocation; $r = 4$--12; all five lengths; $M = 200$ null fits per cell). Fields as in Table~\ref{tab:robust}. $b$: Smith exponent implied by the length elasticity of the error SD of plain strip means (Ag3.0), $b = -2 \times$ elasticity.}
\label{tab:elas-fields}
\small
\begin{tabular}{llccccccc}
\toprule
 & & \multicolumn{3}{c}{Length elasticity} & \multicolumn{3}{c}{Replication elasticity} & \\
\cmidrule(lr){3-5}\cmidrule(lr){6-8}
Layout & Field & Ag3.0 & Ag3.AR1 & Px3.AR1.plot & Ag3.0 & Ag3.AR1 & Px3.AR1.plot & $b$ \\
\midrule
Long strip & Base     & $-0.34$ & $-0.45$ & $-0.34$ & $-0.63$ & $-0.86$ & $-0.63$ & 0.68 \\
           & Range 10 & $-0.40$ & $-0.46$ & $-0.41$ & $-0.65$ & $-0.82$ & $-0.67$ & 0.79 \\
           & Range 60 & $-0.28$ & $-0.43$ & $-0.26$ & $-0.72$ & $-0.91$ & $-0.71$ & 0.55 \\
           & Nugget   & $-0.37$ & $-0.45$ & $-0.39$ & $-0.72$ & $-0.85$ & $-0.73$ & 0.74 \\
Stacked    & Base     & $-0.34$ & $-0.44$ & $-0.25$ & $-0.70$ & $-1.03$ & $-0.77$ & 0.68 \\
           & Range 10 & $-0.39$ & $-0.46$ & $-0.38$ & $-0.68$ & $-1.06$ & $-0.72$ & 0.79 \\
           & Range 60 & $-0.31$ & $-0.40$ & $-0.20$ & $-0.67$ & $-1.13$ & $-0.72$ & 0.63 \\
           & Nugget   & $-0.36$ & $-0.45$ & $-0.39$ & $-0.65$ & $-1.00$ & $-0.65$ & 0.73 \\
\bottomrule
\end{tabular}
\end{table}

\subsection{When do pixel-level analyses pay off?}
\label{sec:pixel}

\begin{figure}[H]
    \centering
    \includegraphics[width=\textwidth]{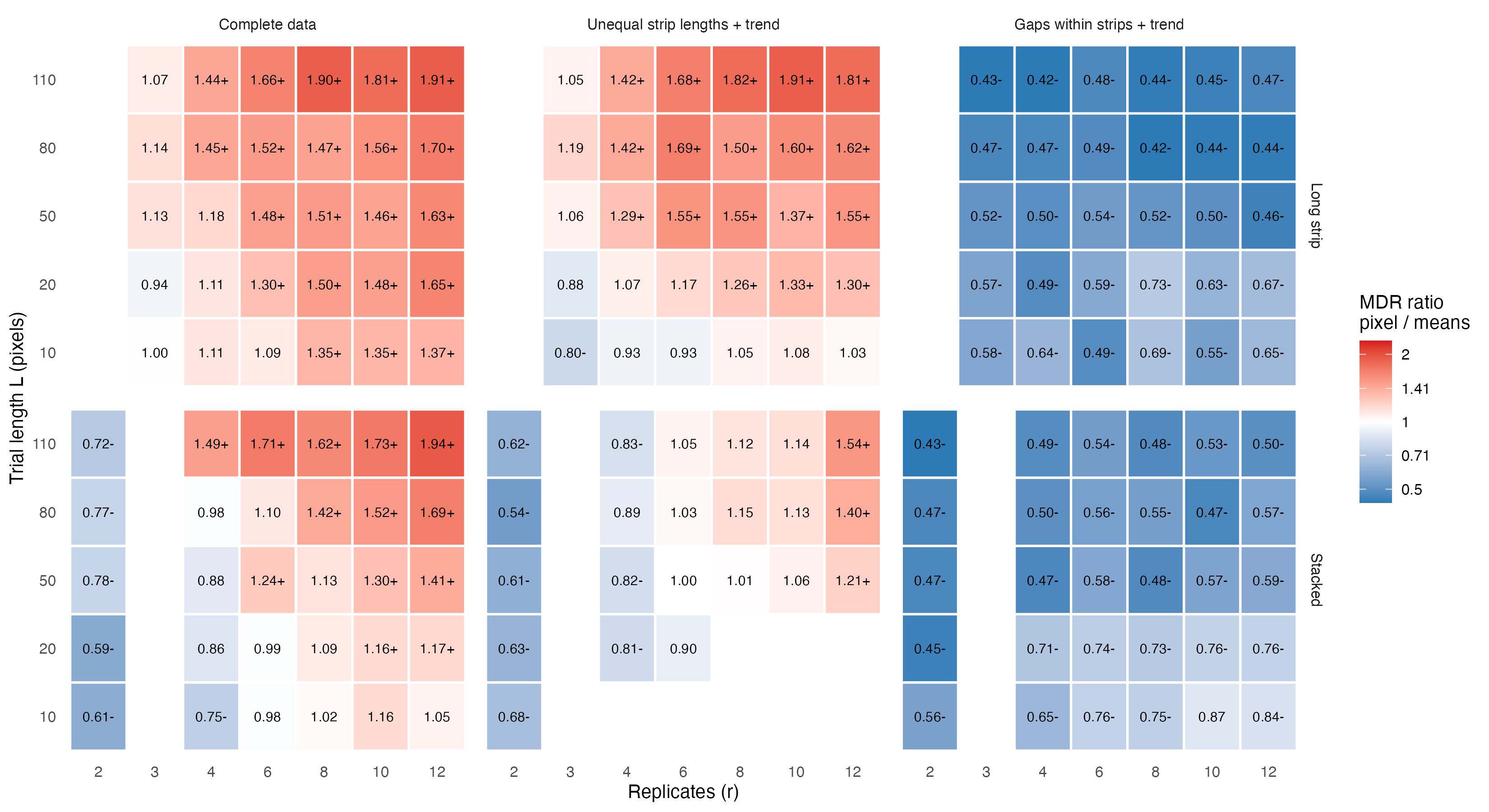}
    \caption{With complete data, strip means detect smaller effects than pixels; with gaps within strips, the pixel model detects effects about half as large. Ratio of the minimum detectable response of the three-column pixel model with a strip effect (Px3.AR1.plot) to that of three-column strip means with an AR1 residual (Ag3.AR1), by trial length and replication, under randomised allocation ($M = 100$ per cell, null and alternative pooled). Blue: pixels detect smaller effects; red: means detect smaller effects. A minus or plus sign marks a cell whose 95\% bootstrap interval excludes 1. Blank cells: the aggregated spatial model converged in too few fits.}
    \label{fig:pixel}
\end{figure}

The main grid favoured aggregation by construction: the response was uniform, the strips were complete and of equal length, and the pixel models used only the centre column of each strip. The follow-up simulations removed these features one at a time (Section~\ref{sec:runs}; Figure~\ref{fig:pixel}; Table~\ref{tab:pixel}).

\paragraph{Complete data.} Using all three pixel columns improved both kinds of analysis. Averaging all three columns lowered the MDR of the aggregated spatial analysis by a median 14--19\% (median ratio of centre-column to three-column MDR, Ag.AR1/Ag3.AR1, 1.16--1.24 over both layouts with complete data and unequal strip lengths, randomised), and fitting all three columns lowered the MDR of the pixel-level spatial model by a quarter at the representative long-strip point (ratio to Ag3.AR1 1.27 against 1.71 for the centre column). With all three columns, Px3.AR1.plot was valid on long strips under randomisation (4.5\%); the centre-column version had been strongly conservative. Strip-plot means nevertheless remained the more precise analysis. On randomised long strips the MDR of Px3.AR1.plot was a median 1.45 times that of Ag3.AR1 (0.94--1.91 over 30 cells; worse in 21 cells, better in none), and the gap widened with both length and replication. On stacked layouts the two were closer (median 1.11, 0.59--1.94), and the pixel model was the more precise at $r = 2$ (0.59--0.78), where the aggregated spatial model has only a few strip means from which to estimate its correlation.

Fitting the covariance family of the simulated field itself, an isotropic exponential, did not close the gap at the representative point (MDR ratio 1.16), and its test was liberal (Type I error 19.8\%). The pixel models estimated the spatial parameters poorly. The centre-column AR1 model estimated the correlation between adjacent strips at 0.19 (median over trials), against 0.90 in the simulated field, and in an example trial the exponential model estimated a correlation of 0.76--0.83 between neighbouring pixels (true 0.97) and a spatial variance of 72{,}000--91{,}000 (true 250{,}000). The trial footprint ($36 \times 50$ pixels) is smaller than the practical range of the field ($\approx 90$ pixels), so within one trial the correlated variation probably looks much like a smooth trend, and its range and variance cannot be separated well. The standard errors of a pixel model, and the weights it gives to individual pixels, depend on these estimates, whereas strip means remove the within-strip structure before any covariance is estimated. For complete strips and a uniform response, the strip means already carry nearly all the information about the treatment contrast.

\paragraph{Unequal strip lengths.} A diagonal paddock edge combined with an along-strip trend changed little on long strips: strip means remained more precise except for the shortest trials ($L = 10$). On stacked layouts, where the edge removed whole plots, the aggregated spatial models often failed to converge (Ag.AR1 53\%, Ag3.AR1 72\% of fits) and Ag3.AR1 was liberal (8.7\%). Px3.AR1.plot, by contrast, converged in every fit, was valid (5.3\%) and matched the precision of the means (median ratio 1.00).

\paragraph{Gaps within strips.} When each plot lost a contiguous block of pixels, the ranking reversed. Px3.AR1.plot had a smaller MDR than Ag3.AR1 in all 30 long-strip cells (median ratio 0.49, range 0.42--0.73) and in 29 of 30 stacked cells (0.56, 0.43--0.87), with valid tests under randomisation (3.4\% and 5.1\%). The advantage grew with strip length, from about 0.6 at $L = 10$--20 to 0.45 at $L \geq 80$ on long strips. At the same time the aggregated spatial analysis became liberal (Ag3.AR1 6.5\% on long strips and 12.7\% on stacked layouts), while the non-spatial analysis of means stayed valid (Ag3.0 4.7\% and 5.6\%) but lost precision. With gaps, each strip mean averages a different section of the strip, so along-strip variation, including the trend, no longer cancels between strips; the pixel model compares neighbouring strips at the same position along the strip and is not affected.

\begin{table}[H]
\centering
\caption{Pixel-level against aggregated analysis, three pixel columns, all design cells (30 per layout; $M = 100$). Ratio: median MDR of the model relative to Ag3.AR1 in the same cell, with the number of cells in which the bootstrap interval excludes 1 in favour of (better) or against (worse) the model. Type I: pooled over cells (\%).}
\label{tab:pixel}
\small
\begin{tabular}{lllccccc}
\toprule
 & & & \multicolumn{3}{c}{Px3.AR1.plot} & \multicolumn{2}{c}{Type I (randomised)} \\
\cmidrule(lr){4-6}\cmidrule(lr){7-8}
Layout & Data & Allocation & Ratio & Better & Worse & Ag3.AR1 & Px3.AR1.plot \\
\midrule
Long strip & Complete        & Randomised & 1.45 & 0  & 21 & 1.5  & 4.5 \\
           &                 & Systematic & 1.65 & 0  & 26 &      &     \\
           & Unequal lengths & Randomised & 1.31 & 1  & 18 & 2.0  & 4.5 \\
           & Gaps            & Randomised & 0.49 & 30 & 0  & 6.5  & 3.4 \\
           &                 & Systematic & 0.60 & 30 & 0  &      &     \\
Stacked    & Complete        & Randomised & 1.11 & 6  & 13 & 6.8  & 3.7 \\
           & Unequal lengths & Randomised & 1.00 & 8  & 3  & 8.7  & 5.3 \\
           & Gaps            & Randomised & 0.56 & 29 & 0  & 12.7 & 5.1 \\
           &                 & Systematic & 0.73 & 19 & 3  &      &     \\
\bottomrule
\end{tabular}
\end{table}

\paragraph{Allocation.} The pixel-level spatial model depended on randomisation. Under systematic allocation Px3.AR1.plot was liberal in every data situation (8.8--22.5\%), against 3.4--5.3\% under randomisation, and its precision advantage with gaps was smaller (median ratio 0.60 on long strips and 0.73 on stacked layouts).

\paragraph{Response maps.} When the response varied continuously across the trial, pixel-level methods recovered the response map better than PE means (Table~\ref{tab:maps}). Over all long-strip cells, GWR with a 6-pixel bandwidth had the lowest map RMSE (1.01 under randomisation), followed by the pixel PE model PEpx3.fAR1.d.plot (1.15), PE means PEag3.0.d (1.23) and a single global slope (1.24). The pixel PE model beat PE means in 90\% of paired trials, and GWR in 80\%. Systematic allocation improved the GWR map (RMSE 0.79 against 1.01; better than PE means in 95\% of trials), consistent with earlier findings that systematic designs suit mapping \parencite{Alesso2021Design, Rakshit2020Novel}. \textcite{cao2024optimal} found a clear advantage for systematic designs mainly under a quadratic response and no significant difference under a linear one; here the response was linear in nitrogen but varied continuously in space, and the systematic advantage held for both bandwidths. Two limits apply. The design-based bandwidth $(\tau - 1)d/2 = 3$\,pixels gave a noisier map than 6\,pixels (RMSE 1.20 against 1.01). And short or lightly replicated trials ($L \leq 20$ with $r \leq 4$) carried too little information for any map: a single global slope was the best estimate in those cells.

\begin{table}[H]
\centering
\caption{Response-map RMSE (kg grain per kg N) for a spatially varying response $\beta(s)$ (mean 6, SD 1.8, exponential range 40 pixels), averaged over all 30 long-strip cells ($M = 100$ per cell); three pixel columns. Better than PEag3.0.d: proportion of paired trials in which the method's map RMSE was lower.}
\label{tab:maps}
\small
\begin{tabular}{lcccc}
\toprule
 & \multicolumn{2}{c}{Randomised} & \multicolumn{2}{c}{Systematic} \\
\cmidrule(lr){2-3}\cmidrule(lr){4-5}
Method & Map RMSE & Better than PEag3.0.d & Map RMSE & Better than PEag3.0.d \\
\midrule
Global slope (Ag3.AR1)          & 1.24 & 51\% & 1.24 & 43\% \\
PE means, $K = 5$ (PEag3.0.d)   & 1.23 & --   & 1.19 & --   \\
Pixel PE, $K = 5$ (PEpx3.fAR1.d.plot) & 1.15 & 90\% & 1.15 & 79\% \\
GWR, bandwidth 3 pixels         & 1.20 & 52\% & 1.05 & 68\% \\
GWR, bandwidth 6 pixels         & 1.01 & 80\% & 0.79 & 95\% \\
\bottomrule
\end{tabular}
\end{table}

\subsection{When the response varies across pseudo-environments}
\label{sec:pe}

\begin{figure}[H]
    \centering
    \includegraphics[width=\textwidth]{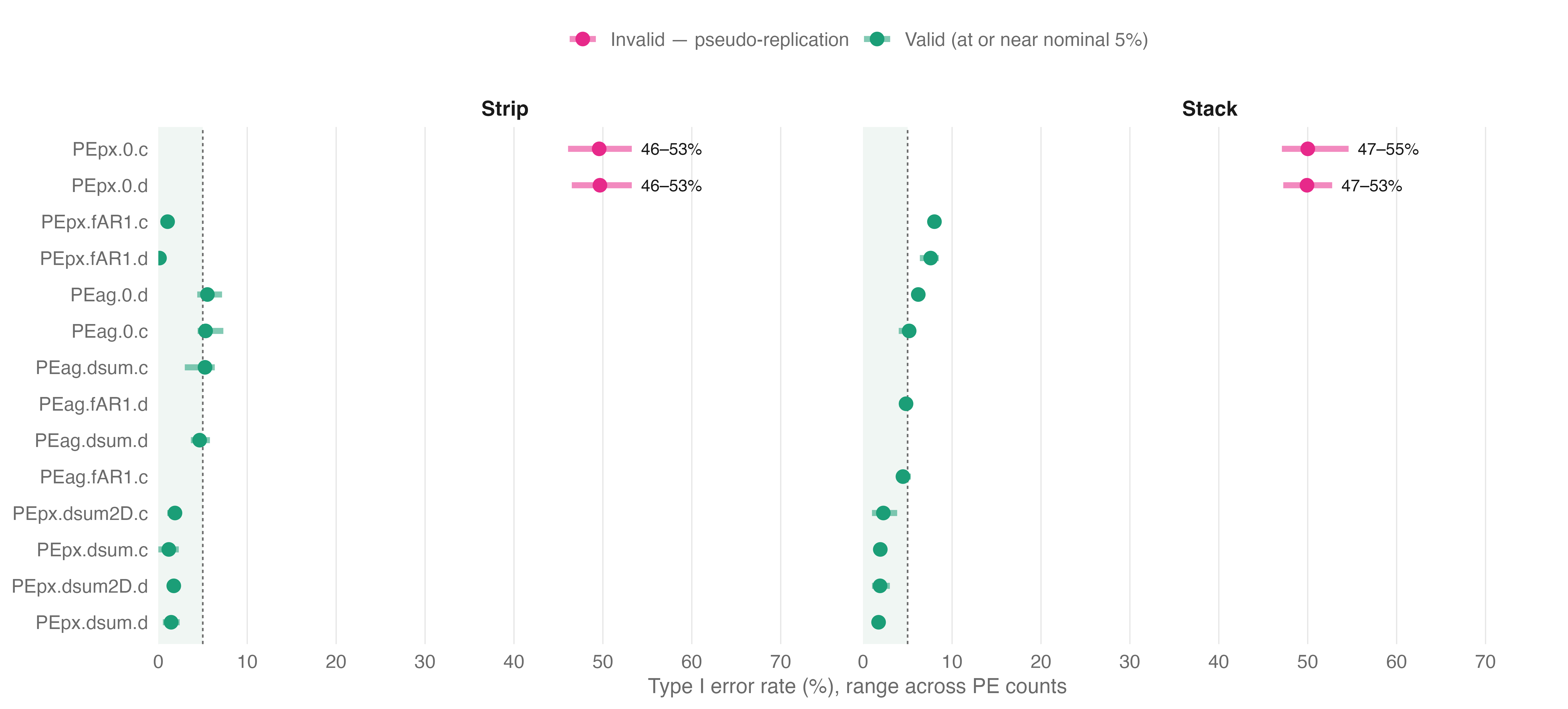}
    \caption{Type I error rate of every Scenario B PE-aware model under the null hypothesis (true treatment effect $= 0$; $M = 100$; equal-size PEs; centre pixel column). The bar spans the range across PE counts and the dot marks the mean; the shaded band and dashed line mark the nominal 5\% level. The pixel non-spatial models PEpx.0.d and PEpx.0.c are invalid through pseudo-replication; all models with a spatial residual, and the aggregated non-spatial models, are at or near the nominal level. PEag.dsum.d/c are absent for stacked layouts, where they are structurally inestimable, and PEag.fAR1.d/c exist for stacked layouts only.}
    \label{fig:scenB-typeI}
\end{figure}

\paragraph{Valid PE-aware analyses.} The calibration step was repeated for the PE-aware suite under the null ($\beta_k = 0$ in every PE, $M = 100$, equal-size PEs; Figure~\ref{fig:scenB-typeI}). The pixel non-spatial models rejected a null rate effect in 46--53\% of PE tests on strips and 47--55\% on stacked layouts, at every PE count; adding PE to the model does not remove pseudo-replication. Aggregation does: PEag.0.d and PEag.0.c sat at 4.4--7.3\% on strips and 4.0--6.5\% on stacked layouts. Every model with a spatial residual controlled Type I error on strips (0.0--6.3\%; pixel-level models $\leq 2.5\%$). On stacked layouts the per-PE residual models (1.0--3.8\%) and the aggregated field-wide model PEag.fAR1.* (3.8--5.5\%) were valid, whereas the pixel-level field-wide model PEpx.fAR1.* was slightly liberal (6.4--8.7\%). As in Scenario~A, the cause was the missing strip term: in the Scenario~B whole-plot run ($M = 300$), adding a random strip effect brought PEpx.fAR1.d from 6.4--9.3\% to 2.7--5.2\% per PE, and its \TPE test from 5.5--14.4\% to 1.2--6.0\%, at some cost in per-PE power (85.7--91.7\% against 93.6--94.8\%).

\begin{figure}[H]
    \centering
    \includegraphics[width=\textwidth]{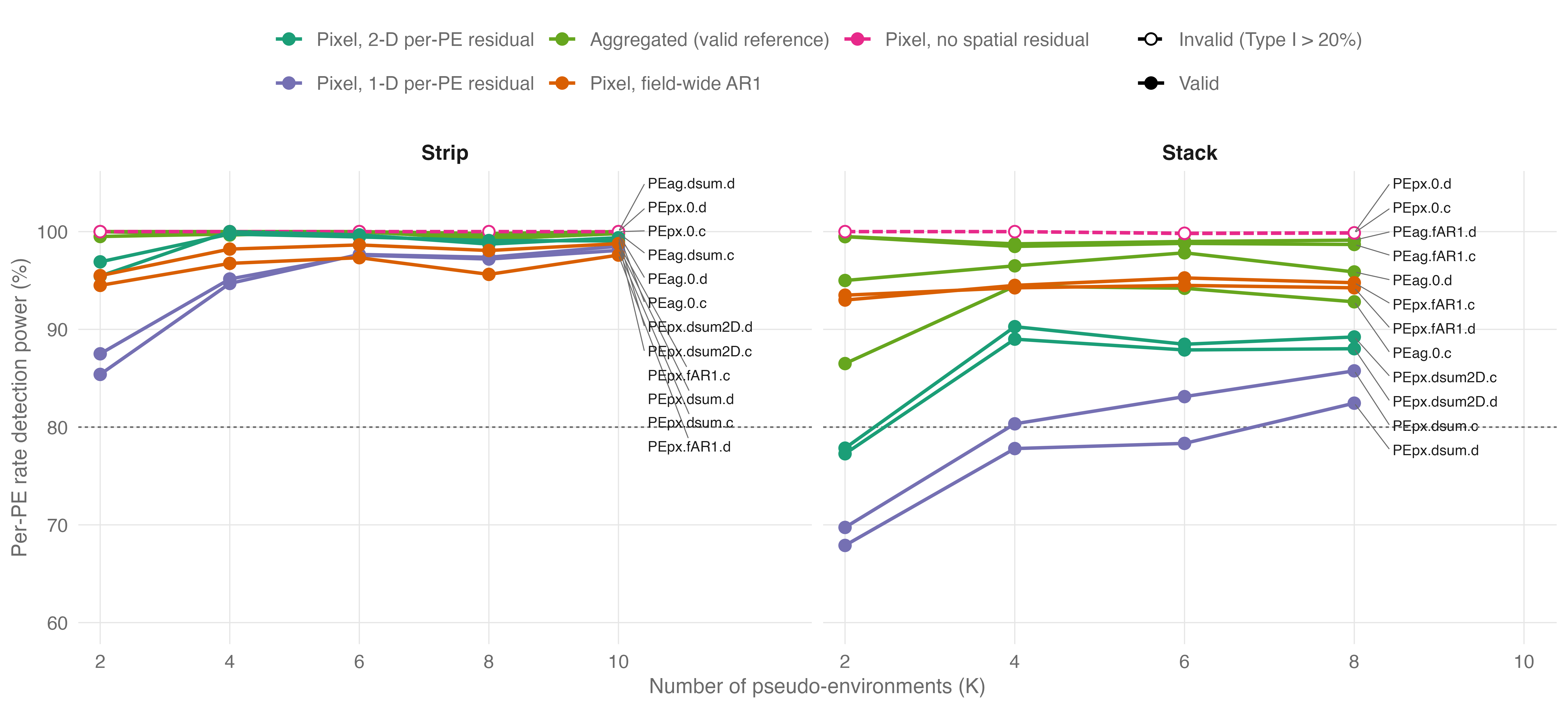}
    \caption{Per-PE rate detection power against PE count for the Scenario B models at the representative design point ($L = 50$ strips, $L = 20$ stacked, $r = 6$; $M = 100$; equal-size PEs; centre pixel column; dashed line at 80\% power). Series drawn with dashed lines and hollow markers are those shown in Figure~\ref{fig:scenB-typeI} to have invalid Type I error, so their power cannot be used for model selection.}
    \label{fig:scenB-power-K}
\end{figure}

\paragraph{How many PEs can a trial support?} Splitting the trial into more PEs gives each PE fewer observations, yet among the valid models per-PE rate detection power did not decline as PEs multiplied (Figure~\ref{fig:scenB-power-K}). On strips, the aggregated models reached 96--99\% per-PE power at every PE count (PEag.0.d 98.1--99.1\%), the pixel-level field-wide model 93.3--95.5\% (PEpx.fAR1.d), and the one-dimensional per-PE residual 87.0--95.3\% (PEpx.dsum.d), with its lowest power at $K = 2$. On stacked layouts the aggregated field-wide model gave the highest power (PEag.fAR1.d 97.7--99.2\%), and the pixel-level per-PE residual models were the weakest valid option (72.5--84.7\% for PEpx.dsum.d). Each PE's contrast is estimated mainly from comparisons between neighbouring strips within that PE, and these remain available however short the PE becomes. \TPE interaction power was $\geq 94\%$ for every valid PE-aware model on strips at $L = 50$ and 99.2--100\% for PEag.fAR1.* on stacked layouts (Appendix~\ref{app:figures}).

\paragraph{Which PE analysis is most precise?} Power was saturated here as well, so the PE models were compared on per-PE MDR in the PE follow-up run (three pixel columns; Table~\ref{tab:pe-pixel}). PE means without a spatial term (PEag3.0.d, the three-column form of PEag.0.d) were the least precise option in every setting. With complete data on long strips, PE means with a field-wide AR1 residual (PEag3.fAR1.d) detected per-PE effects 43\% smaller (median ratio 0.57, all 12 cells), and the pixel PE model 36\% smaller (0.64). On stacked layouts the pixel PE model was the most precise (0.67 against 0.77 for the aggregated spatial model). With gaps, the pixel PE model was the most precise on both layouts (0.54 on long strips, 0.59 on stacked layouts; all 12 cells), and the advantage of the aggregated spatial model shrank (0.87 and 0.94). The pixel advantage did not grow with the number of PEs: with gaps, its ratio moved from 0.49 at $K = 2$ to 0.57 at $K = 6$ on long strips, and from 0.58 to 0.63 on stacked layouts. A two-dimensional per-PE residual (PEpx3.dsum2D.d.plot) was slightly less precise than the field-wide residual throughout (0.60--0.73). Taken together, the pixel PE model with a strip effect and a field-wide residual was the most precise option in three of the four layout and data combinations, and within 12\% of the best in the fourth.

\begin{table}[H]
\centering
\caption{PE-aware analyses, three pixel columns, randomised allocation (PE follow-up run: long strip $L \in \{50, 110\}$, stacked $L \in \{20, 50\}$, $r \in \{4, 8\}$, $K \in \{2, 4, 6\}$; 12 cells per layout and data situation; $M = 100$). Ratio: median per-PE MDR relative to PEag3.0.d, with the number of cells in which the model was significantly more precise. Type I: per-PE rejection rate under the null; FWER: probability that at least one PE test rejects, without and with Bonferroni correction (all \%).}
\label{tab:pe-pixel}
\small
\begin{tabular}{lllcccccc}
\toprule
Layout & Data & Model & Ratio & Better & Type I & FWER & Bonferroni & RAD \\
\midrule
Long strip & Complete & PEag3.0.d         & 1.00 & --    & 5.7 & 19.2 & 5.1  & 7.0 \\
           &          & PEag3.fAR1.d      & 0.57 & 12/12 & 1.4 & 5.0  & 1.2  & 3.9 \\
           &          & PEpx3.fAR1.d.plot & 0.64 & 12/12 & 6.6 & 19.4 & 8.8  & 4.4 \\
           & Gaps     & PEag3.0.d         & 1.00 & --    & 5.7 & 19.4 & 6.1  & 10.4 \\
           &          & PEag3.fAR1.d      & 0.87 & 10/12 & 7.2 & 25.3 & 10.7 & 8.6 \\
           &          & PEpx3.fAR1.d.plot & 0.54 & 12/12 & 8.2 & 22.6 & 9.7  & 5.7 \\
Stacked    & Complete & PEag3.0.d         & 1.00 & --    & 6.2 & 19.2 & 5.7  & 10.6 \\
           &          & PEag3.fAR1.d      & 0.77 & 9/12  & 6.2 & 17.5 & 7.1  & 8.0 \\
           &          & PEpx3.fAR1.d.plot & 0.67 & 12/12 & 5.4 & 16.9 & 7.4  & 7.0 \\
           & Gaps     & PEag3.0.d         & 1.00 & --    & 6.5 & 20.8 & 6.7  & 13.9 \\
           &          & PEag3.fAR1.d      & 0.94 & 5/12  & 10.7 & 30.4 & 15.9 & 12.6 \\
           &          & PEpx3.fAR1.d.plot & 0.59 & 12/12 & 6.6 & 19.5 & 8.0  & 8.5 \\
\bottomrule
\multicolumn{9}{p{15.5cm}}{\footnotesize PEag3.0.d and PEag3.fAR1.d use strip $\times$ PE means over all three columns; PEag3.fAR1.d has an \texttt{ar1(column):ar1(segment)} residual on both layouts, unlike the stacked-only PEag.fAR1.d of the main grid. RAD: mean per-PE relative absolute deviation of $\hat\Delta_{120}$ (\%).}
\end{tabular}
\end{table}

The spatial PE models came at a cost in calibration. The pixel PE models were slightly liberal under randomisation (per-PE Type I error 5.4--8.2\% for PEpx3.fAR1.d.plot) and more so under systematic allocation (13.0--19.8\%). With gaps, the aggregated spatial model was also liberal (7.2\% and 10.7\%).

\paragraph{The cost of more PEs is multiplicity.} Each PE-level test holds its own level only approximately, and a trial with $K$ PEs makes $K$ tests. In the Scenario~B whole-plot run, the probability of a false positive in at least one PE for PEag.0.d rose from 10.3\% at $K = 2$ to 33.7\% at $K = 8$ and 38.0\% at $K = 10$ on strips, and from 9.7\% to 34.3\% at $K = 8$ on stacked layouts; a Bonferroni correction (testing each PE at $0.05/K$) held it at 4.3--6.7\% on strips and 3.0--9.3\% on stacked layouts. In the PE follow-up run (Table~\ref{tab:pe-pixel}), uncorrected FWER was 17--30\% for every model except the conservative PEag3.fAR1.d on complete long strips, and Bonferroni reduced it to 5--16\%. A trial designed to support PE-specific recommendations should either correct for multiplicity or first test the \TPE interaction and examine individual PEs only when it is significant.

\paragraph{Per-PE effects must be tested conditionally.}
\label{sec:test-form}
The incremental run told a different story about the pixel-level field-wide model: under \asreml{}'s default incremental Wald chi-square test, its per-PE power fell from 99.5\% at $K = 2$ to 39.0\% at $K = 10$ on strips (PEpx.fAR1.c: 99.7\% to 50.9\%), which would suggest that a field-wide residual cannot cope with PE-varying responses. Under conditional $F$-tests the same model held 93.3--95.5\%. The estimates were unbiased in both runs (mean ratio of estimated to true PE response 0.99--1.01 in every PE at $L = 50$). An incremental test assesses each PE's rate term after adjusting only for the terms that precede it; under a field-wide residual the PE rate terms are correlated, so only the last PE receives a properly adjusted test. At $K = 10$ and $L = 50$, the last PE was detected in 100\% of trials and the first nine in only 13--62\%; on the same 20 data sets (200 PE tests), the incremental test detected 38.5\% of PE effects and the conditional test 97.5\%. The form of the Wald test is part of the analysis, and for PE-specific terms conditional $F$-tests must be requested explicitly.

\paragraph{Delineating PEs: boundary placement dominates PE size.} The simulations above divide the trial into equal PEs, but real productivity zones are unequal. Crossing PE size inequality with PE count showed no effect of inequality itself on the non-spatial or field-wide models, which converged in 94--100\% of fits whether PEs were equal or differed by a factor of about six. Per-PE dsum residuals failed only when the smallest PE was reduced to a single row (0.0\% convergence; 98.6\% at two rows). Letting each PE choose its own residual structure (PEpx.dsumAd.d) recovered 97.8\% convergence in those cells, with per-PE power of 98.9\% and Type I error of 3.2\% (Appendix~\ref{app:pe}).

\begin{figure}[H]
    \centering
    \includegraphics[width=\textwidth]{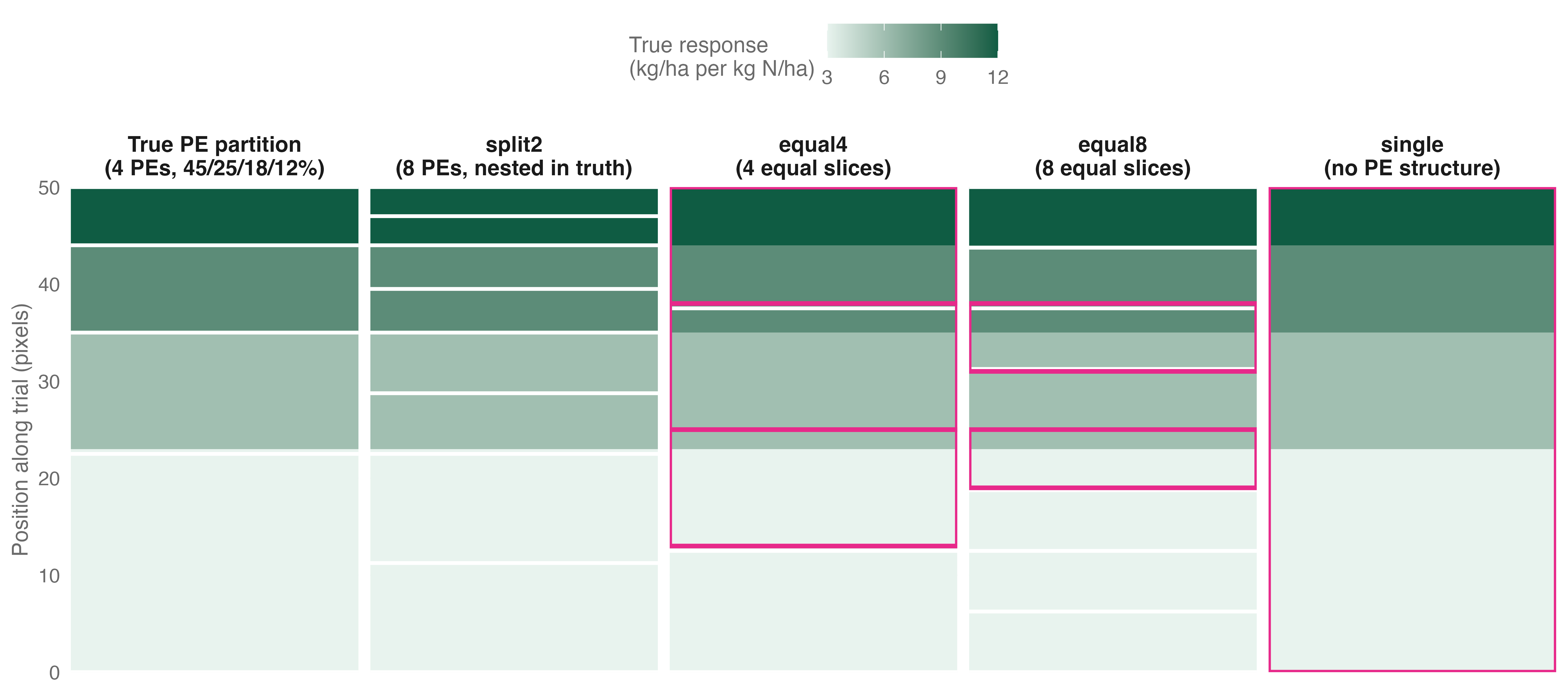}
    \caption{The five analysis partitions applied to one true field. Fill shows the true, spatially varying rate response; the true PEs are unequal, as real productivity zones are. Pink outlines mark analysis PEs that span more than one true PE, so that their estimated response is an average of different true responses. Subdividing within the true boundaries (split2) produces no such PE; imposing equal slices (equal4, equal8) produces several.}
    \label{fig:pe-partition-schematic}
\end{figure}

\begin{figure}[H]
    \centering
    \includegraphics[width=\textwidth]{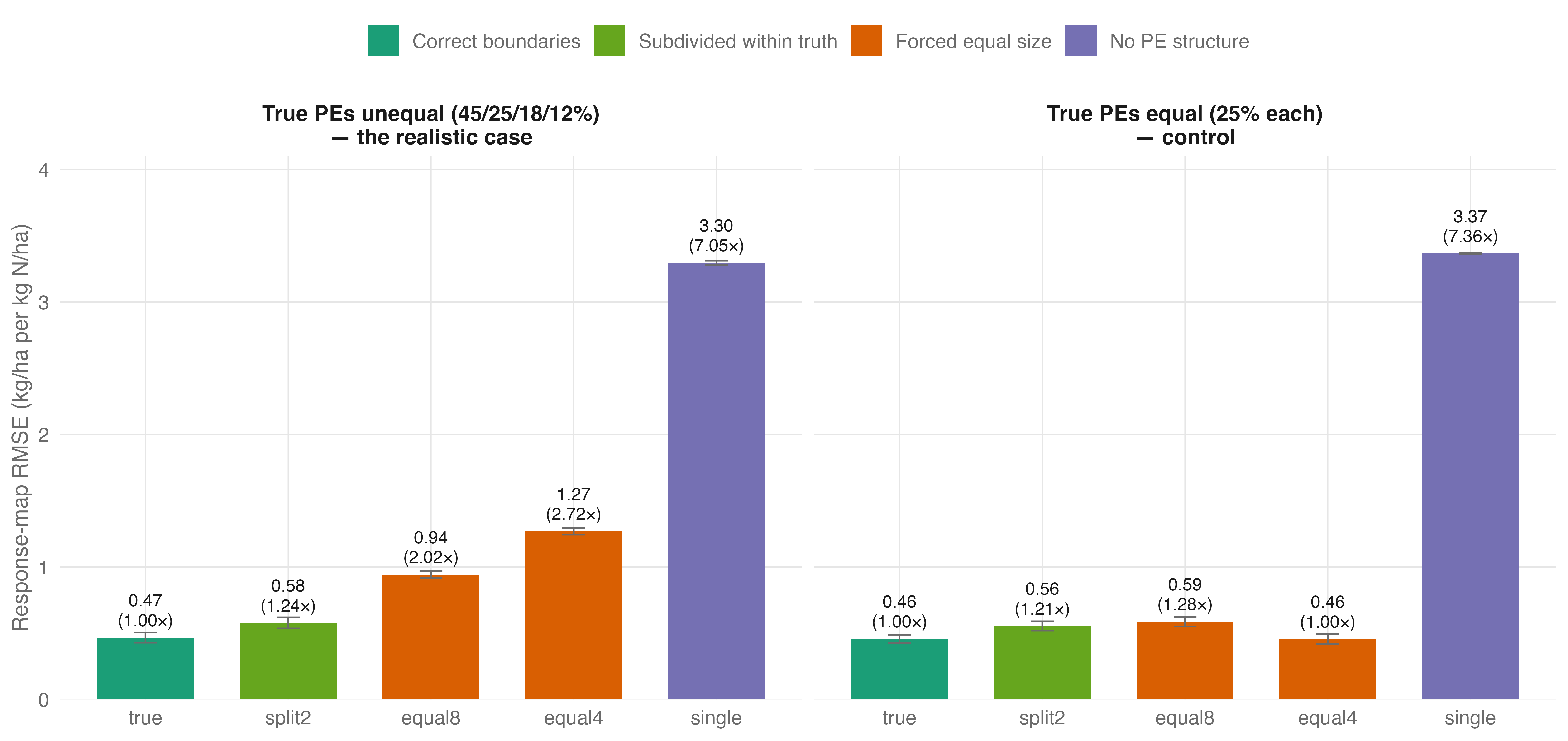}
    \caption{Root mean squared error of the recovered response map under each analysis partition, for model PEpx.dsumAd.d ($M = 100$; error bars are 95\% Monte Carlo intervals). Left: with unequal true PEs, forcing equal-size analysis PEs costs 2.7 times the RMSE of the correct partition, while subdividing within true boundaries costs 1.24 times. Right: when the true PEs are themselves equal, the equal-slice partition coincides with the truth and performs identically, isolating boundary misalignment rather than size inequality as the source of the penalty.}
    \label{fig:pe-partition-rmse}
\end{figure}

To separate boundary placement, resolution and size balance, one true field was re-analysed under five analysis partitions (Figure~\ref{fig:pe-partition-schematic}). The field had four unequal true PEs with response coefficients of 3, 6, 9 and 12\,kg grain per kg N, and each partition was scored by the root mean squared error of the recovered response map. Imposing four equal slices cost 2.7 times the map RMSE of the correct four-PE partition, and ignoring PE structure altogether 7.1 times, whereas subdividing each true PE in half cost only 1.24 times (Figure~\ref{fig:pe-partition-rmse}). When the true PEs were themselves equal, the equal-slice partition coincided with the truth and performed identically (0.457 against 0.458). The penalty comes from boundary misalignment, not from size inequality. It is a bias rather than a variance penalty: an analysis PE that straddles two true PEs estimates an average of two different responses, and more replication cannot remove it. With non-monotone response patterns the equal-slice penalty widened to 4.6--7.0 times, while nested subdivision still cost 1.25--1.27 times.

\subsection{A design table for strip trials}
\label{sec:framework}

Table~\ref{tab:decision} brings the results together as a sequence of decisions to be made before a trial is sown. The first rows fix the analysis, because a design can be judged only through a valid one; the middle rows set the design; and the last rows apply when the trial is to support PE-specific recommendations or response maps.

\begin{table}[H]
\centering
\caption{Design table for large strip trials. MDR: minimum detectable response at 80\% power.}
\label{tab:decision}
\small
\singlespacing
\begin{tabular}{c>{\raggedright\arraybackslash}p{2.6cm}>{\raggedright\arraybackslash}p{6.3cm}>{\raggedright\arraybackslash}p{6.3cm}}
\toprule
Step & Element & Recommendation & Evidence \\
\midrule
1 & Validity & Never treat yield pixels as independent replicates. Analyse strip-plot means, or include a random strip effect in any pixel-level model & Px.0 32--40\% Type I; PEpx.0.* 46--55\%; a strip effect restores validity \\
2 & Test & Conditional Wald $F$-tests (\texttt{denDF = "default"}, \texttt{ssType = "conditional"}) & Chi-square: Ag.0 8.7\% vs 5.2\%; incremental per-PE power 39\% vs 95\% \\
3 & Effect size & Decide the smallest gain worth detecting, and read $L$ and $r$ off Figure~\ref{fig:mdr} & Long strip, $L = 50$, $r = 4$: 4.9\% of mean yield (Ag3.AR1), 8.4\% (Ag3.0) \\
4 & Replication & The strongest lever: lay out as many replicates as the paddock allows; at least $r = 4$ on long strips & MDR $\propto r^{-0.60}$ to $r^{-1.02}$; more, shorter replicates won all 66 matched-area comparisons \\
5 & Length & Run strips the full paddock length; when choosing between longer strips and more replicates, choose replicates & MDR $\propto L^{-0.24}$ to $L^{-0.45}$; RAD keeps falling to $L = 110$ \\
6 & Allocation & Randomise for testing; use a systematic layout only when a response map is the main aim & Systematic: Ag.0 5.9--12.6\%, Px3.AR1.plot 8.8--22.5\% Type I; GWR map RMSE 0.79 vs 1.01 \\
7 & Layout & Prefer long strips; stack only to buy replicates in a short paddock, then randomise and aim for $r \geq 8$ if strip means are analysed spatially & Same plots: long strip better or equal; same footprint: stacked MDR 17--23\% smaller (randomised) \\
8 & Data & Use every pixel column of each strip, for means and for pixel models & Three-column means: MDR a median 14--19\% smaller \\
9 & Analysis, complete data & Strip-plot means with an AR1 residual (Ag3.AR1); on stacked layouts with $r < 8$, Ag3.0 or a pixel model with a strip effect & Pixel/means MDR 1.45 (long strip), 1.11 (stacked) \\
10 & Analysis, gaps in strips & Pixel-level AR1 model with a strip effect (Px3.AR1.plot), randomised trials only & MDR halved (0.49--0.56); Ag3.AR1 6.5--12.7\% Type I \\
11 & PE analysis & Default: pixel-level PE model with a strip effect and a field-wide AR1 residual (PEpx3.fAR1.d.plot), as in \textcite{stefanova2023}, with per-PE tests corrected for multiplicity (Step 12); avoid PE means without a spatial term & Per-PE MDR 0.54--0.67 of PEag3.0.d; most precise in 3 of 4 settings; PE means with AR1 marginally better on complete long strips (0.57); per-PE Type I 5.4--8.2\%, Bonferroni FWER 7.4--9.7\% \\
12 & Number of PEs & Per-PE power holds as PEs multiply, but correct for multiplicity or test \TPE first & Any-PE false positive 10--38\% uncorrected; Bonferroni 3--16\% \\
13 & PE delineation & Delineate on the defining covariates and leave PEs unequal; subdivide only within boundaries; ensure $\geq 2$ rows per PE or relax that PE's residual & Equal slices cost 2.7--7.0$\times$ the map RMSE; nested subdivision 1.24--1.27$\times$ \\
14 & Response maps & Pixel-level GWR or pixel PE models, bandwidth of about two strip widths, systematic layout, $L \geq 50$ or $r \geq 6$ & GWR RMSE 0.79--1.01 vs 1.19--1.23 for PE means; global slope best for short trials \\
\bottomrule
\end{tabular}
\end{table}

\section{Discussion}
\label{sec:discussion}

\subsection{What makes a strip trial informative}

Ranked by how much they change what a trial can detect, the design elements studied here fall in the following order. Replication comes first: each doubling of replicates cut the detectable response to 49--66\% of its previous value, and replication won every comparison against length at equal plot area. The analysis comes second, and it is a lever of the same size: an autoregressive residual on strip means lowered the detectable response by a median 38--49\%, and when strips contained gaps a pixel-level model halved it again. Length comes third: it helped throughout the range studied but at a diminishing rate, because consecutive pixels along a strip are correlated. Layout and allocation matter mainly through validity rather than precision; they decide which analyses can be trusted.

Power, the usual design criterion, could not have revealed this order. At the simulated response, power was saturated for almost every valid design, and on power alone the designs would have appeared nearly interchangeable, with only Type I error separating the analyses. The minimum detectable response restores the comparison. It agreed with simulated power at smaller effects to within 4 percentage points for the aggregated analyses, so for these analyses it can be read as what a trial will detect. In these terms, a randomised long-strip trial with four replicates of 500\,m strips, analysed with strip means of all three pixel columns and an AR1 residual (Ag3.AR1), detects a 0--120\,kg\,N gain of about 5\% of mean yield; doubling the replicates more than halves that (2.2\% at $r = 8$), whereas more than doubling the strip length, to 1.1\,km, does not (2.9\% at $L = 110$).

\subsection{Four design intuitions, reconsidered}

\emph{``Longer strips are better''.} True, with diminishing returns. Precision kept improving up to the longest strips simulated (1.1\,km), with no plateau in accuracy, but the gain per unit length was smaller than independent observations would give, and smaller than the gain from replication. Where paddock length is fixed and costs nothing extra, strips should run its full length; length is not, however, a substitute for replication.

\emph{``More replicates are better''.} True, and the strongest lever, as the simulations of \textcite{Alesso2019Experimental} also suggested. Replication improves precision at about or above the rate expected for independent units, and each replicate also adds error degrees of freedom, which matters most at the low replication typical of OFE. It also brings stacked layouts to a valid spatial analysis ($r \geq 8$). The practical minimum on long strips is four replicates: at three, even the best analysis detects only gains of 6--14\% of mean yield, depending on length.

\emph{``Randomisation is better than systematic allocation''.} True for testing, false for mapping. Randomisation did not improve the precision of the analyses of strip means, and systematic allocation was marginally more precise for them on long strips, consistent with its more regular spatial coverage of treatment levels and with \textcite{Alesso2019Experimental}, who found that randomisation increased the uncertainty of treatment estimates. Its value is insurance: under systematic allocation the classical analysis of strip means was liberal on both layouts, with or without an explicit trend, and so were pixel-level spatial models fitted to all pixel columns. Only a spatial residual on strip means stayed valid without randomisation. When the aim is a response map, the balance reverses: systematic allocation gave a GWR map error a fifth lower than randomisation, in line with \textcite{Alesso2021Design}, and did so for a linear response, for which \textcite{cao2024optimal} had found no significant difference. The choice of allocation should follow the main inferential purpose of the trial.

\emph{``Long strips are better than stacked layouts''.} True plot for plot, false paddock for paddock. With the same number and size of plots, the long-strip arrangement was almost never less precise. On a fixed paddock footprint, the stacked layout buys twice the replicates, and for every analysis under randomisation this outweighed the arrangement penalty. The stacked layout's weakness is validity: it depends on randomisation, on a strip term in any pixel model, and on enough replicates for the spatial analysis of strip means.

\subsection{Matching the analysis to the data}

The comparison between aggregated and pixel-level analysis does not have a single answer. With complete strips of equal length and a uniform response, the strip means already contain nearly all the information about the treatment contrast: along-strip variation is averaged over the same section of every strip and cancels between treatments, and an AR1 residual on the means accounts for the variation across strips. In that situation a pixel-level model can at best match the means, and in practice it fell short, because its spatial parameters have to be estimated within a trial that is smaller than the correlation range of the field. This held even when the pixel model used the covariance family of the simulated field. The disadvantage depends on the size of the trial relative to the correlation range of the field. In the heterogeneity runs, the ratio of the error SD of the centre-column pixel model to that of Ag.AR1 fell from 1.53 at range parameter 30 to 1.24 at range 10 on long strips (1.67 to 1.46 on stacked layouts), and rose to 1.59 (1.78) at range 60. Where a trial spans several correlation ranges, as large trials on fields with short-range yield variation may, pixel-level models are likely to perform better than they did here; this deserves testing on fields and trials of realistic size.

Real yield-monitor data are often incomplete. Harvester turns, blockages, overlaps and outlier screening remove records in blocks \parencite{Sudduth2007Yield, Lyle2014Postprocessing, Vega2019Protocol}, and paddock edges cut strips to different lengths. When each strip lost a section, its mean no longer described the same part of the paddock as the means of its neighbours, the along-strip variation stopped cancelling, and the detectable response of the means doubled. The pixel-level model compares neighbouring strips at the same position along the strip and was largely unaffected. In this situation the pixel-level analysis with a strip effect halved the detectable response and kept a valid test, whereas the spatial analysis of strip means became liberal. These results support the pixel-level, pseudo-environment approach of \textcite{stefanova2023} for real strip-trial data, with two requirements that the simulations make explicit: the model must include a random effect for the strip, and the treatments must be randomised. Using every pixel column of each strip, rather than a single transect, improved both kinds of analysis.

Pixel-level data also carry information that strip means cannot: the location of the response. When the response varied within the trial, GWR and pixel PE models recovered the response map more accurately than PE means, with the best results from GWR with a bandwidth of about two strip widths under systematic allocation. The design-based bandwidth was too narrow in our setting, and short or lightly replicated trials carried too little information for any map, so mapping is a reason to plan a longer, well-replicated, systematic trial rather than a reason to analyse an ordinary comparative trial differently.

The form of the test is part of the analysis. Chi-square tests made the classical analysis of strip means look liberal, and incremental tests made a valid field-wide PE model look powerless. Conditional $F$-tests should be requested and reported. Among the PE random-effect structures, diag(zone) is preferred to corh(zone): it converged 18--21 percentage points more often on stacked trials at a modest cost in power (Appendix~\ref{app:pe}).

\subsection{Zoned trials}

For trials meant to support PE-specific decisions, the results are favourable for power but call for care with inference. Per-PE power did not decline as PEs multiplied, because each PE's contrast rests on neighbouring strips within it. The choice of PE model matters more than the number of PEs. The better spatial PE model detected per-PE effects 33--46\% smaller than PE means without a spatial term, which would otherwise be the obvious choice. The pixel PE model with a strip effect was the most precise on stacked layouts and whenever the data contained gaps. On complete long strips it was within 12\% of the best model, PE means with a field-wide AR1, which was marginally more precise and conservative. Because large strip trials are usually zoned and their yield data are rarely complete, these results support the pixel-level pseudo-environment approach of \textcite{stefanova2023} as the default analysis for such trials. The advantage of the pixel PE model did not grow with the number of PEs, so it rests on the data situation rather than on zoning as such.

What does grow with the number of PEs is the chance of a spurious PE-specific effect: without correction, about one trial in five to one in three reports a response in some PE that has none. A multiplicity correction, or a \TPE test that gates the PE-level tests, should be part of the design, particularly for the pixel PE models, whose per-PE tests were slightly liberal. PE delineation should follow the covariates that define the zones: misplaced boundaries cost far more than unequal sizes, because a PE that straddles two true zones estimates a blend of their responses, which no amount of replication can undo.

\subsection{Implications for OFE practice}

\emph{Before the trial}, the grower and adviser should decide the smallest yield gain worth detecting and choose the design from Figure~\ref{fig:mdr}: as many replicates as the paddock allows (at least four on long strips), strips running the full paddock length, randomised allocation for a comparative trial, and a stacked layout only where it is the way to obtain the replication needed. A systematic layout is appropriate when the main aim is a response map. If PE-specific recommendations are wanted, the PEs should be drawn from soil, elevation or historical-yield information, without adjusting boundaries to equalise areas.

\emph{At the analysis stage}, the analyst should use every pixel column of each strip and include a random strip effect in any pixel-level model. With complete data, strip-plot means with an AR1 residual are a sound default; when the cleaned yield data have gaps or uneven strip lengths, a pixel-level AR1 model with a strip effect is the better choice. For zone-specific questions, a pixel-level PE model with a strip effect and a field-wide AR1 residual is the recommended default, PE means without a spatial term should be avoided, and PE-level tests should be conditional $F$-tests corrected for multiplicity.

\emph{For software and service providers}, automated OFE tools should not fit models to yield-monitor pixels without a term for the strip, should check how much of each strip is missing before choosing between aggregated and pixel-level analysis, should use conditional $F$-tests for zone-specific effects, and could report the detectable response of a planned trial as a design aid.

\subsection{Limitations and scope}
\label{sec:limitations}

\emph{Simulated conditions.} The main grid used one realisation of a stationary exponential field and the centre pixel column of each strip. The supplementary fields (three realisations each of five heterogeneity structures) confirmed the validity results and the ranking of the analyses at the representative design points. The follow-up runs repeated the design comparisons, and the comparison of aggregated and pixel-level analyses, with all three pixel columns across the full grid. The length and replication elasticities were re-estimated on four heterogeneity structures; the ordering of the two levers held in all of them, but the size of the length elasticity depended on the correlation range, so the values in Table~\ref{tab:elasticity} are specific to the base field. The simulated correlation range (practical range about 900\,m) exceeded the trial dimensions, which handicaps the estimation of pixel-level covariance; trials spanning several correlation ranges were examined only at the representative design points. The gap and trend patterns were stylised: real data gaps are clustered by machinery operations and real trends follow landforms.

The simulated trial footprint is the paddock: the study does not model headlands, non-stationary soil zones, or the choice of where in a larger field to site a trial, and it does not address whether a trial's estimate represents the paddock as a whole when the response varies with location. These questions of trial siting and representativeness require a different simulation design and are left for future work. The response was additive and linear in nitrogen, whereas real responses are typically nonlinear and interact with soil and yield potential. Finally, the recommendations have not yet been applied to observed strip-trial data; applying them to public OFE datasets would show how they play out in practice.

\emph{Design criterion.} The MDR uses the RCBD error degrees of freedom as the denominator degrees of freedom for every analysis; the agreement with simulated power indicates that this is adequate for the aggregated analyses, and the MDR describes the precision that a calibrated test could use rather than the power of a miscalibrated one. The matched-area comparisons treat $L \times r$ as the budget, which measures plot area rather than cost; where strip length is free and replicates are constrained by paddock width or machinery, the practical trade-off differs.

\emph{Inference.} All measures are conditional on model convergence, which ranged from about 50\% to 100\% across the models compared (for example, 53\% for Ag.AR1 on stacked trials with unequal strip lengths); the per-PE residual models PEag.dsum.* could not be fitted to stacked aggregated data at all (Appendix~\ref{app:pe}). A single trial whose chosen model fails to converge requires a stated fallback. The pixel-level models were not perfectly calibrated: they were conservative with the centre column only, slightly liberal in some PE analyses, and liberal under systematic allocation; small-sample adjustments of the denominator degrees of freedom, such as that of \textcite{Kenward1997Small}, were not examined. The Scenario~B null simulation, the whole-plot refits and the two-dimensional and adaptive per-PE residuals were evaluated at one representative design point per layout, and the PE follow-up run at eight design points. The PE-partitioning results treat boundaries as either correct or displaced by construction and do not model the error of estimating boundaries from covariates, so the ``correct partition'' benchmark is an upper bound on what a real zoning exercise can achieve; PEs were also varied only along the trial axis.

\section{Conclusion}
Large strip trials let growers test management options at the scale at which decisions are made, but the rules used to design and analyse them have largely been borrowed from small-plot research. This study tested four design rules and the choice between aggregated and pixel-level analysis by simulation, scoring each design by the smallest response that a valid analysis detects.

Replication is the strongest design lever: it improves precision faster than trial length does, and at equal plot area more, shorter replicates beat fewer, longer ones in every comparison. Longer strips still help, so strips should run the full paddock length. Randomisation does not improve precision but keeps the tests valid, for the classical analysis of strip means and for pixel-level models alike; systematic allocation is preferable only when the aim is a response map. Long strips are at least as precise as stacked layouts plot for plot, but stacking is a legitimate way to buy replicates in a short paddock, provided treatments are randomised. The analysis should match the data: strip-plot means with an AR1 residual for complete data, and pixel-level spatial models with a strip effect when strips contain gaps, as cleaned yield-monitor data often do. With gaps, the pixel-level model halved the detectable response. When responses vary across PEs, a pixel-level PE model with a strip effect is the recommended default: it detected per-PE effects 33--46\% smaller than PE means without a spatial term. Trials retain per-PE power as PEs multiply, and multiplicity must be controlled.

The small-plot intuitions are therefore not wrong but unequal and conditional, and the design table (Table~\ref{tab:decision}) and the detectable-gain chart (Figure~\ref{fig:mdr}) set out those conditions in a form that can be used before a trial is sown. They are intended for growers, advisers and analysts in Australian broadacre OFE: to design trials that can detect the effects that matter to them, and to analyse the resulting data in the way that suits it.

\section*{CRediT authorship contribution statement}
\textbf{Zhanglong Cao:} Conceptualization, Methodology, Software, Formal analysis, Investigation, Data curation, Visualization, Writing -- original draft, Writing -- review \& editing.
\textbf{Suman Rakshit:} Formal analysis, Validation, Writing -- review \& editing.
\textbf{Barbara Kachigunda:} Methodology, Writing -- review \& editing.
\textbf{Julia Easton:} Writing -- review \& editing.
\textbf{Mark Gibberd:} Supervision, Writing -- review \& editing.

\section*{Declaration of competing interest}
The authors declare that they have no known competing financial interests or personal relationships that could have appeared to influence the work reported in this paper.

\section*{Data availability}
The R code used to run the simulations and to summarise the results, and the summary result files, are available from the corresponding author on request.

\section*{Declaration of generative AI and AI-assisted technologies in the writing process}
During the preparation of this work the authors used Claude (Anthropic) to assist with programming, summarising simulation results and editing the text. After using this tool, the authors reviewed and edited the content as needed and take full responsibility for the content of the published article.

\section*{Acknowledgements}
This work was supported by the Grains Research and Development Corporation (GRDC) through the Analytics for the Australian Grains Industry (AAGI) research activity AAGI-CU-RD-OFE.

\printbibliography

\appendix
\section{Model specifications}
\label{app:models}

\begin{table}[H]
\centering
\caption{Scenario A statistical models. ``Aggregated'' data uses the mean yield per strip plot. For strip layouts, Ag.AR1 uses \texttt{id(range):ar1(row)} (a 1-D AR1 across adjacent strip means, with one range level, because aggregation removes the along-strip dimension).}
\label{tab:models-scena}
\begin{tabular}{llll}
\toprule
Model & Data & Spatial residual & Random effects \\
\midrule
Ag.0  & Aggregated & None                          & rep \\
Ag.AR1  & Aggregated & id(range):ar1(row)$^\dagger$  & rep \\
Px.0  & Pixel & None                          & rep \\
Px.AR1  & Pixel & ar1(range):ar1(row)           & rep \\
Px.AR1w & Pixel & id(range):ar1(row)            & rep \\
Px.AR1.plot & Pixel & ar1(range):ar1(row)      & rep $+$ plot \\
\bottomrule
\multicolumn{4}{l}{$^\dagger$Stacked layouts with $r > 2$ use \texttt{ar1(range):ar1(row)}.}
\end{tabular}
\end{table}

\begin{table}[H]
\centering
\caption{Scenario B PE-aware models (main grid). Each model is also fitted in an interaction (\TPE) form, suffixed \texttt{.i} (\texttt{PE + rate + PE:rate} in model notation), used for interaction testing. PEag.dsum.d and PEag.dsum.c are structurally inestimable on stacked aggregated data (0\% convergence).
         $^\dagger$For stacked-replicate layouts, \texttt{zone\_rep} (within-PE replicate index) replaces \texttt{rep};
         for strip layouts, the global \texttt{rep} factor is used directly.
         PEag.fAR1.d and PEag.fAR1.c exist for stacked layouts only.
         $^\ddagger$PEpx.dsum2D.d/c apply a two-dimensional autoregressive process independently within each PE, in contrast to the one-dimensional PEpx.dsum.d/c, which treat adjacent strips as independent.
         $^\S$PEpx.dsumAd.d/c select each PE's residual structure from that PE's own realised geometry using \texttt{dsum(..., levels\,=\,)}: \texttt{ar1(range):ar1(row)} where the PE spans at least two rows and two ranges, and \texttt{ar1(range):id(row)} or \texttt{id(range):ar1(row)} where one dimension is degenerate. The PE follow-up run fitted PEag3.0.d, PEag3.fAR1.d (\texttt{ar1(column):ar1(segment)} on strip $\times$ PE means), PEpx3.fAR1.d.plot and PEpx3.dsum2D.d.plot (both with \texttt{diag(zone):rep + plot}) to all three pixel columns.}
\label{tab:models-scenb}
\small
\begin{tabular}{lllll}
\toprule
Model  & Data       & Spatial residual               & Random effects          & PE structure \\
\midrule
PEag.0.d  & Aggregated & None                           & diag(zone):rep          & at(zone):rate \\
PEag.dsum.d  & Aggregated & dsum(id(range):ar1(row)|zone)  & diag(zone):rep          & as PEag.0.d \\
PEag.dsum.c  & Aggregated & dsum(id(range):ar1(row)|zone)  & corh(zone):rep$^\dagger$ & as PEag.0.d \\
PEag.0.c  & Aggregated & None                           & corh(zone):rep$^\dagger$ & as PEag.0.d \\
PEag.fAR1.d  & Aggregated & ar1(range):id(row) field-wide  & diag(zone):rep          & as PEag.0.d \\
PEag.fAR1.c  & Aggregated & ar1(range):id(row) field-wide  & corh(zone):rep$^\dagger$ & as PEag.0.d \\
PEpx.0.d  & Pixel & None                           & diag(zone):rep          & as PEag.0.d \\
PEpx.dsum.d  & Pixel & dsum(id(range):ar1(row)|zone)  & diag(zone):rep          & as PEag.0.d \\
PEpx.0.c  & Pixel & None                           & corh(zone):rep$^\dagger$ & as PEag.0.d \\
PEpx.dsum.c  & Pixel & dsum(id(range):ar1(row)|zone)  & corh(zone):rep$^\dagger$ & as PEag.0.d \\
PEpx.dsum2D.d & Pixel & dsum(ar1(range):ar1(row)|zone)$^\ddagger$ & diag(zone):rep & as PEag.0.d \\
PEpx.dsum2D.c & Pixel & dsum(ar1(range):ar1(row)|zone)$^\ddagger$ & corh(zone):rep$^\dagger$ & as PEag.0.d \\
PEpx.dsumAd.d & Pixel & dsum, per-PE structure$^\S$    & diag(zone):rep          & as PEag.0.d \\
PEpx.dsumAd.c & Pixel & dsum, per-PE structure$^\S$    & corh(zone):rep$^\dagger$ & as PEag.0.d \\
PEpx.fAR1.d  & Pixel & ar1(range):ar1(row) field-wide & diag(zone):rep          & as PEag.0.d \\
PEpx.fAR1.c  & Pixel & ar1(range):ar1(row) field-wide & corh(zone):rep$^\dagger$ & as PEag.0.d \\
\bottomrule
\end{tabular}
\end{table}

\section{Calibration, power and accuracy}
\label{app:figures}

\begin{figure}[H]
    \centering
    \includegraphics[width=\textwidth]{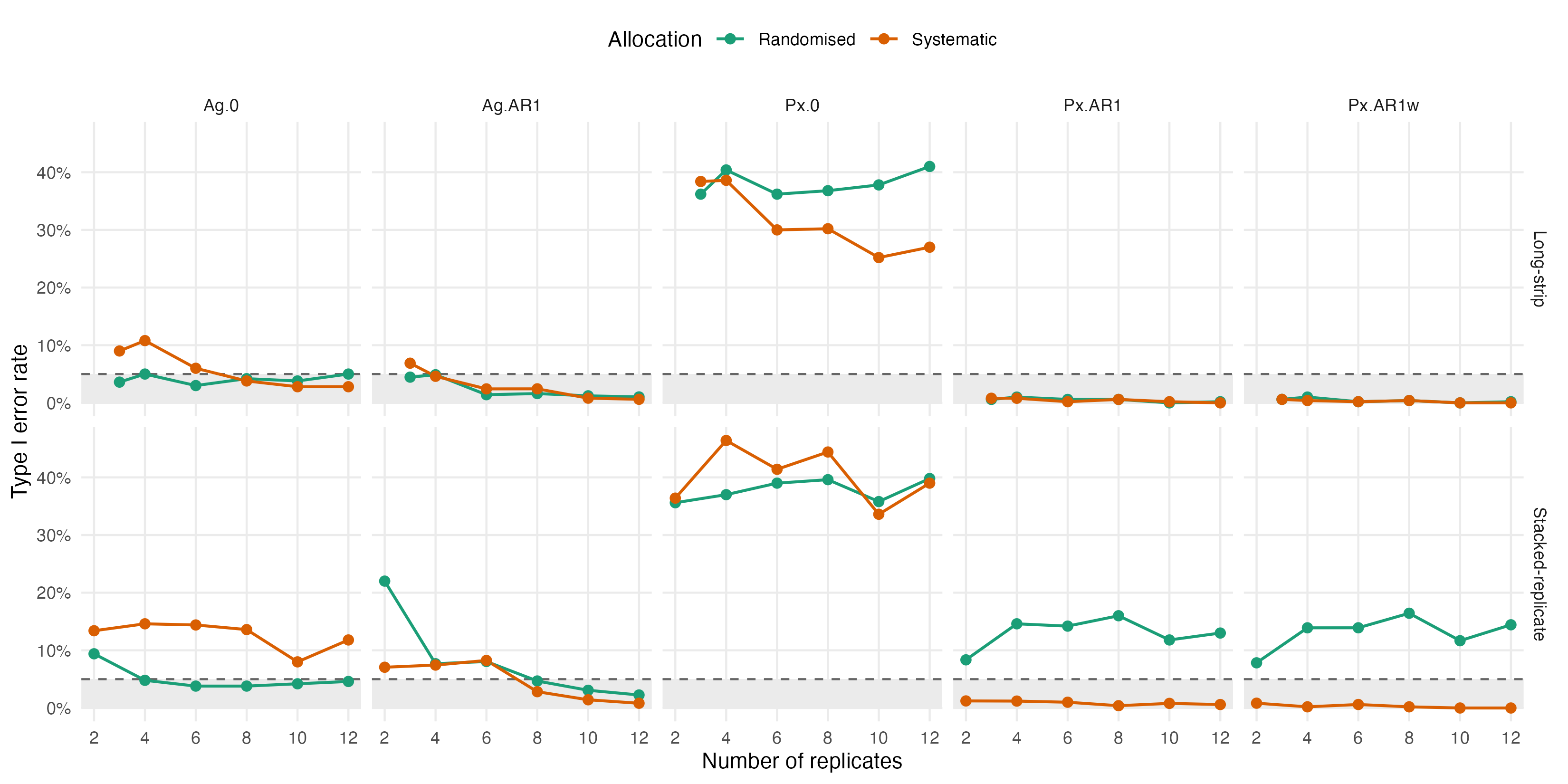}
    \caption{Pixels analysed without a term for the strip give invalid tests on both layouts. Type I error rate against the number of replicates for the Scenario~A models, by layout (rows) and allocation (colour), averaged over trial lengths ($M = 100$ per cell). The shaded band and dashed line mark the nominal 5\% level. Px.0 is invalid on both layouts. On stacked layouts, the pixel spatial models without a plot term are liberal under randomised allocation and Ag.0 is liberal under systematic allocation.}
    \label{fig:typeI}
\end{figure}

\begin{figure}[H]
    \centering
    \includegraphics[width=\textwidth]{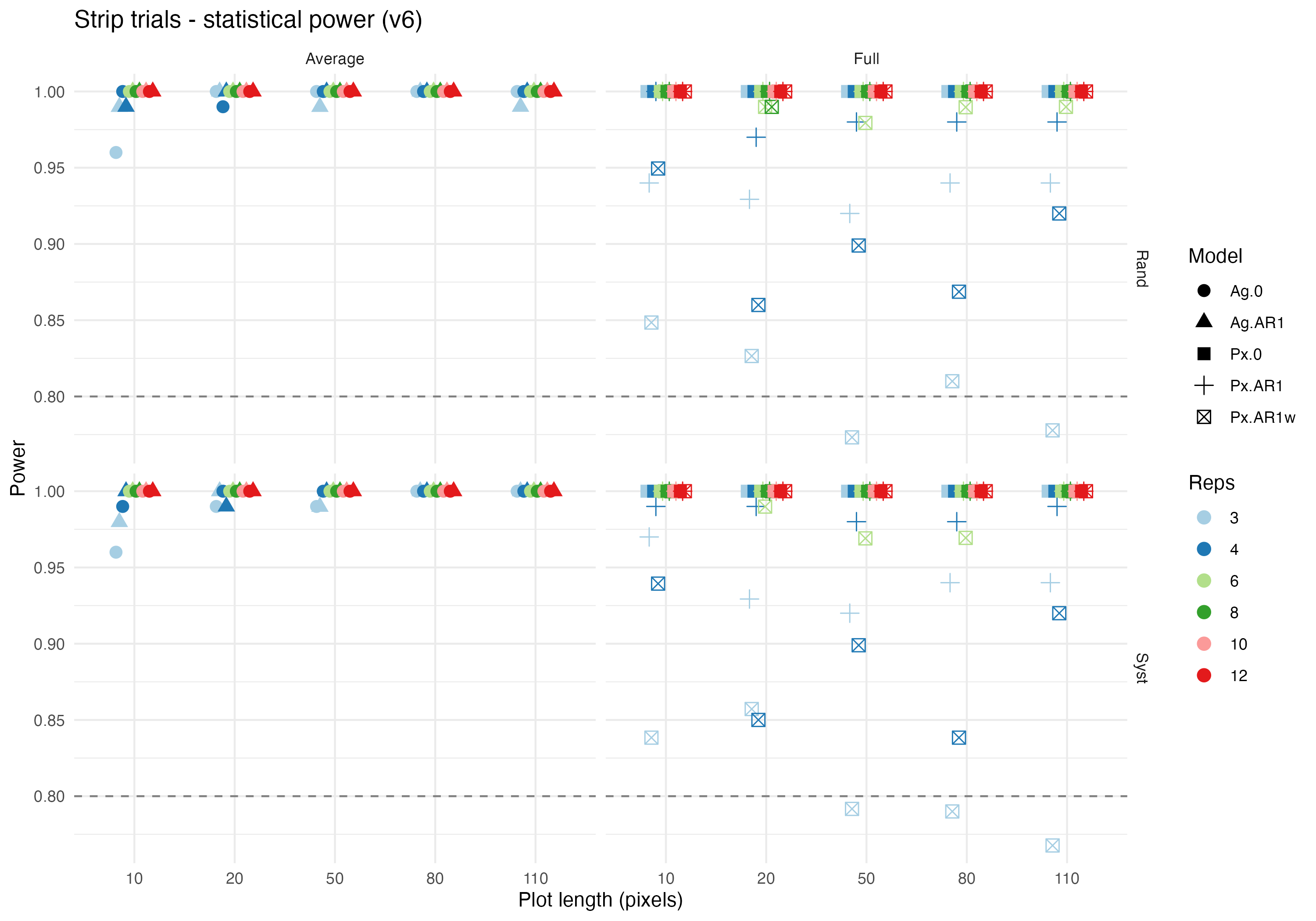}
    \caption{Statistical power of long-strip trials at $\beta = 6$ as a function of trial length (pixels) and number of replicates, faceted by model. Px.0 is shown for completeness but is invalid (Table~\ref{tab:typeI}). Dashed line: 80\% threshold.}
    \label{fig:strip-power}
\end{figure}

\begin{figure}[H]
    \centering
    \includegraphics[width=\textwidth]{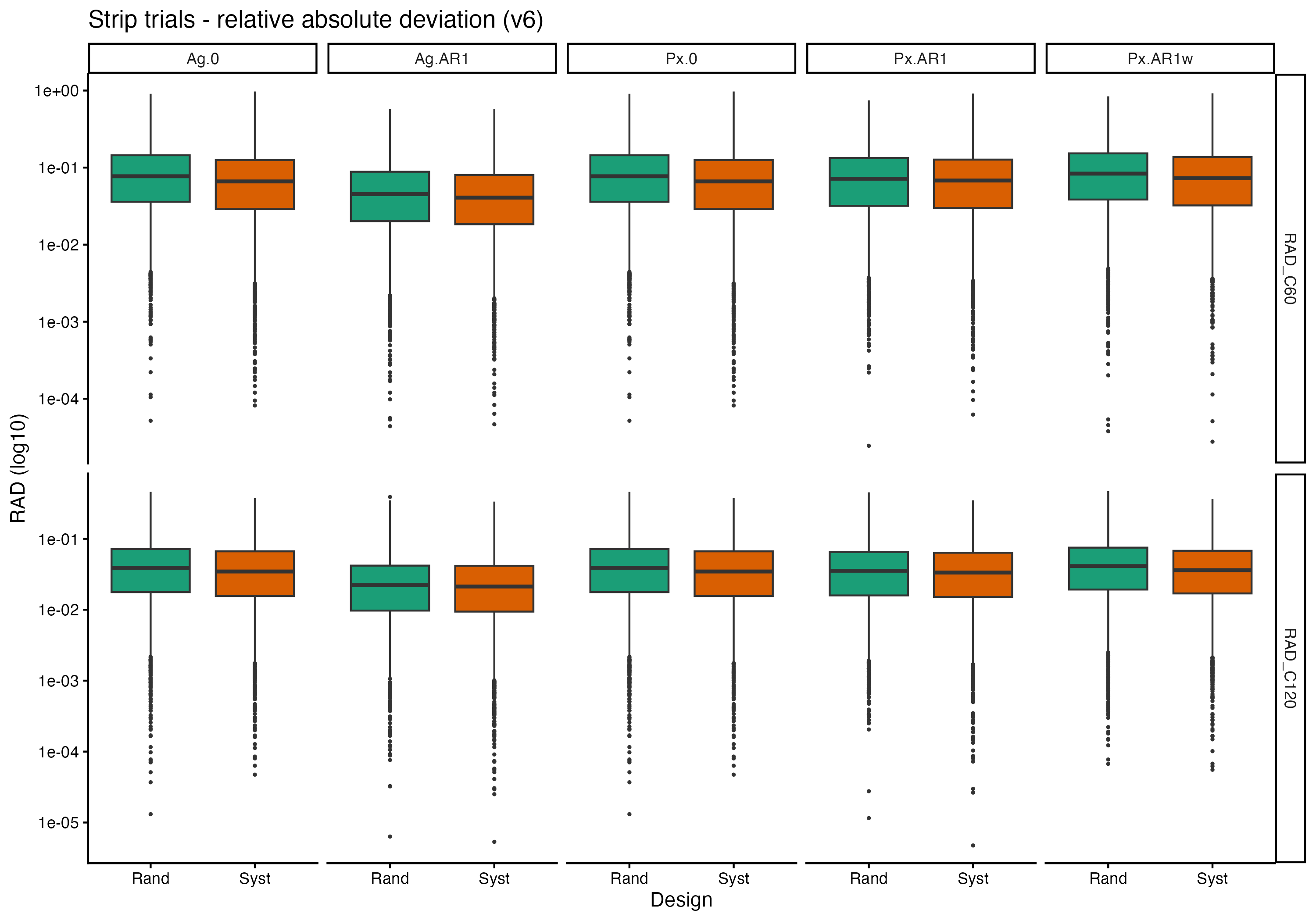}
    \caption{Relative absolute deviation (RAD) of treatment-contrast estimates for long-strip trials (log scale).}
    \label{fig:strip-rad}
\end{figure}

\begin{figure}[H]
    \centering
    \includegraphics[width=\textwidth]{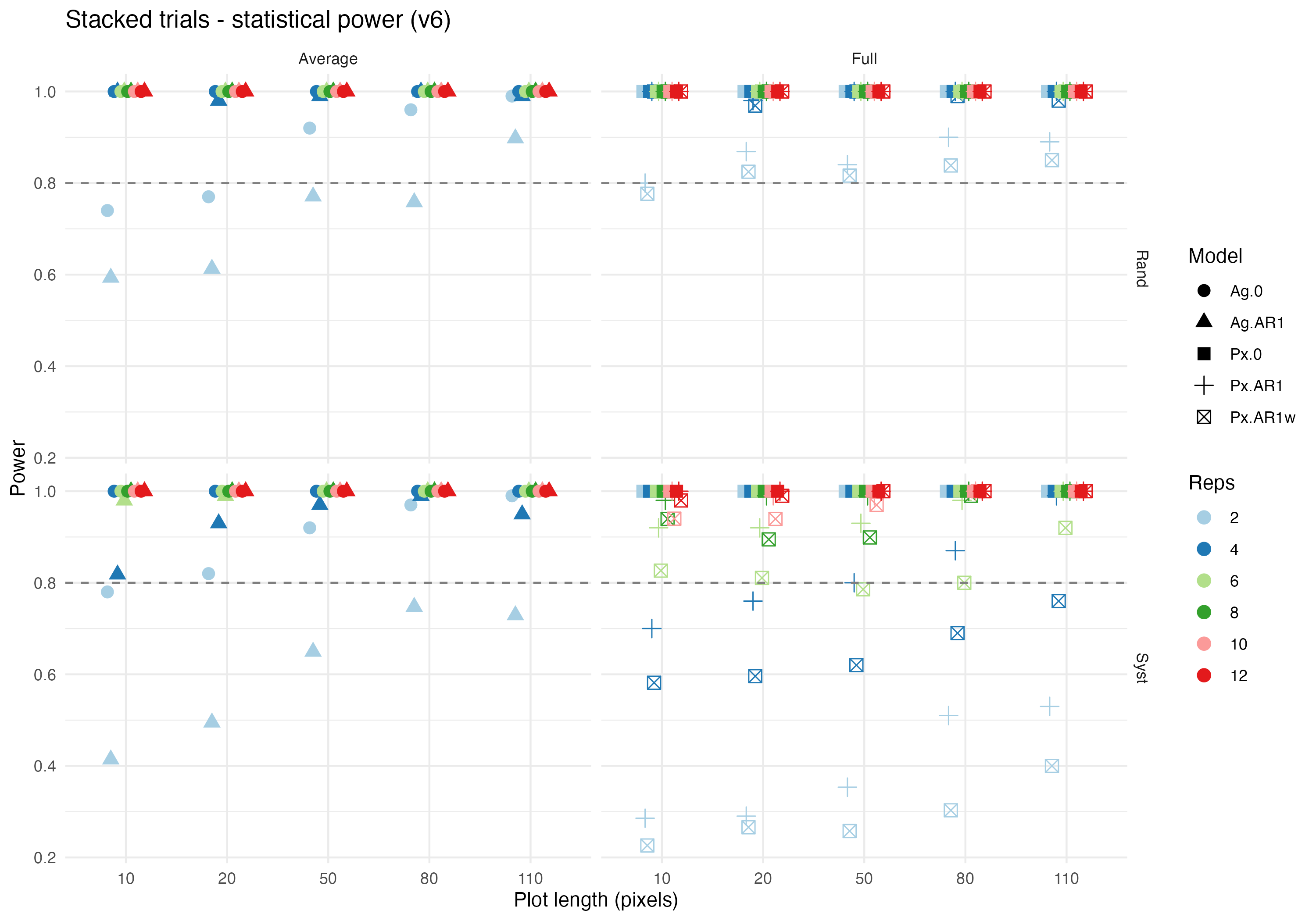}
    \caption{Statistical power of stacked-replicate trials at $\beta = 6$ as a function of replicates and trial length, faceted by model.}
    \label{fig:stack-power}
\end{figure}

\begin{figure}[H]
    \centering
    \includegraphics[width=\textwidth]{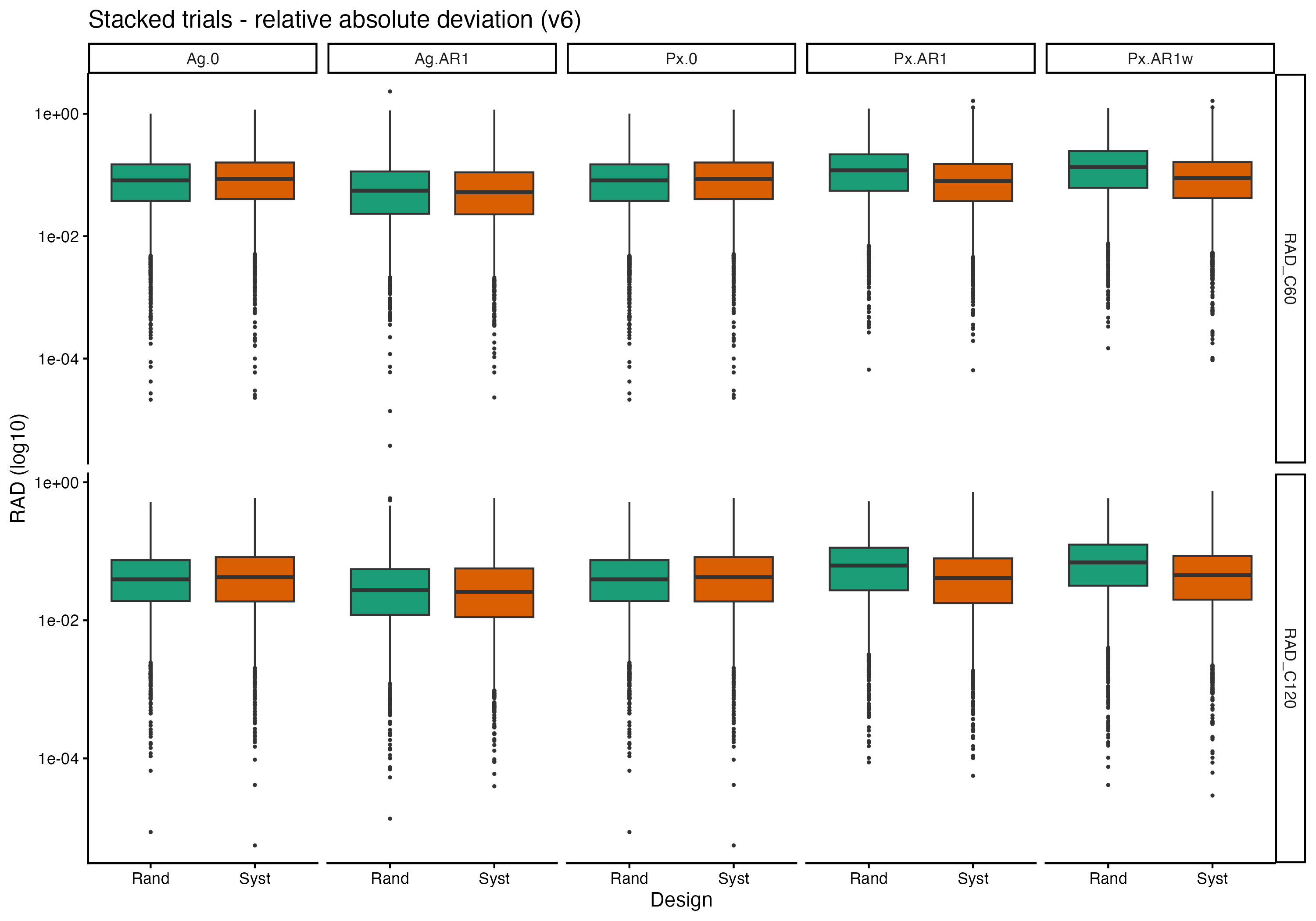}
    \caption{Relative absolute deviation (RAD) for stacked-replicate trials (log scale).}
    \label{fig:stack-rad}
\end{figure}

\begin{figure}[H]
    \centering
    \includegraphics[width=\textwidth]{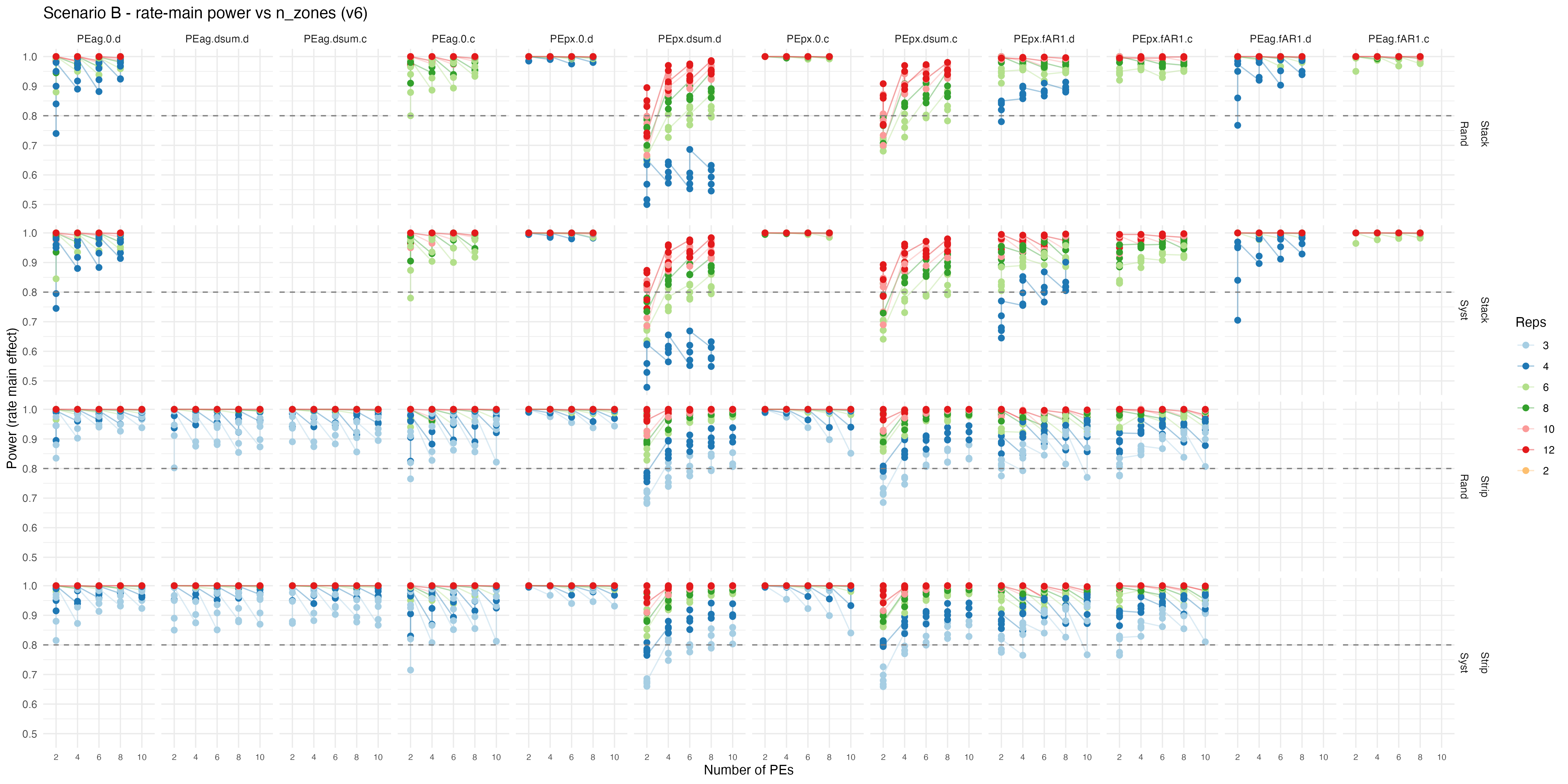}
    \caption{Scenario B: per-PE rate detection power as a function of the number of PEs, by model (columns), layout and allocation (rows), and replication (colour). PEpx.0.d and PEpx.0.c are invalid (Figure~\ref{fig:scenB-typeI}). Dashed line: 80\%.}
    \label{fig:scenB-rate}
\end{figure}

\begin{figure}[H]
    \centering
    \includegraphics[width=\textwidth]{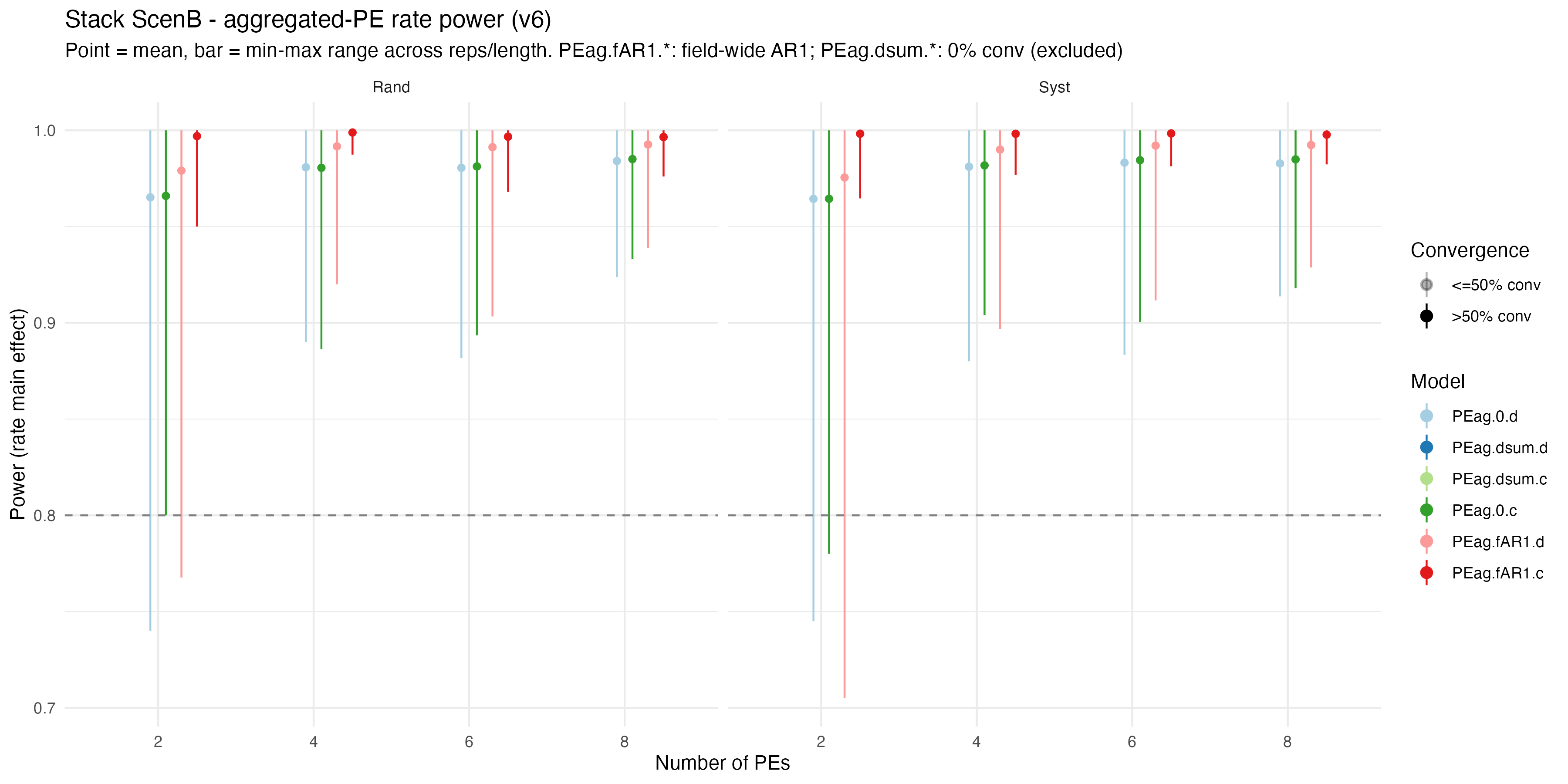}
    \caption{Scenario B (stacked trials): rate detection power for the aggregated-PE models PEag.0.d, PEag.0.c, PEag.fAR1.d and PEag.fAR1.c.}
    \label{fig:scenB-agg-rate}
\end{figure}

\begin{figure}[H]
    \centering
    \includegraphics[width=\textwidth]{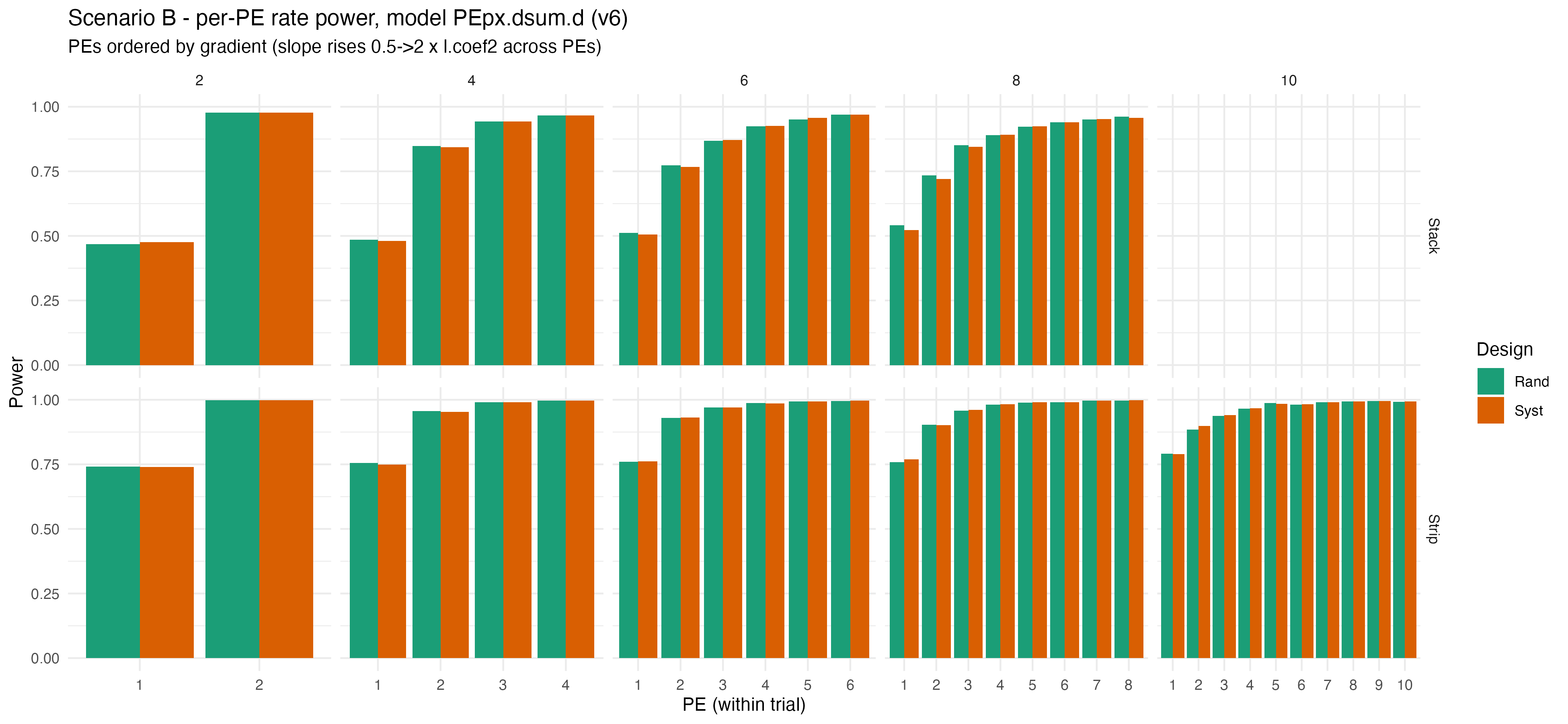}
    \caption{Per-PE rate detection power for model PEpx.dsum.d (pixel, one-dimensional per-PE residual) as a function of PE count, illustrating the variation in detection probability across individual PEs.}
    \label{fig:per_zone_dsum_d}
\end{figure}

\begin{figure}[H]
    \centering
    \includegraphics[width=\textwidth]{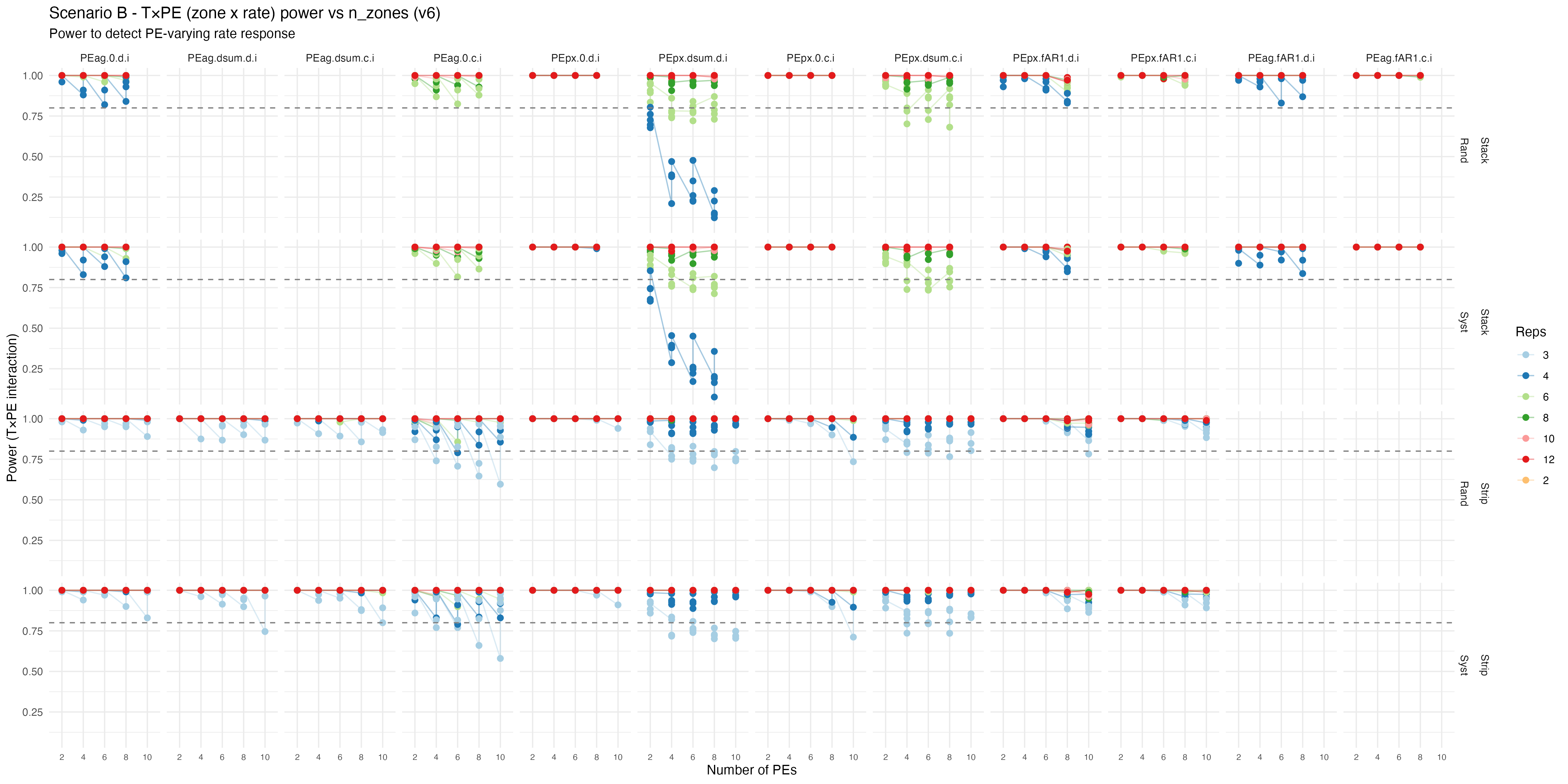}
    \caption{Scenario B: \TPE interaction detection power as a function of number of PEs. Dashed line: 80\%.}
    \label{fig:scenB-gxt}
\end{figure}

\begin{figure}[H]
    \centering
    \includegraphics[width=\textwidth]{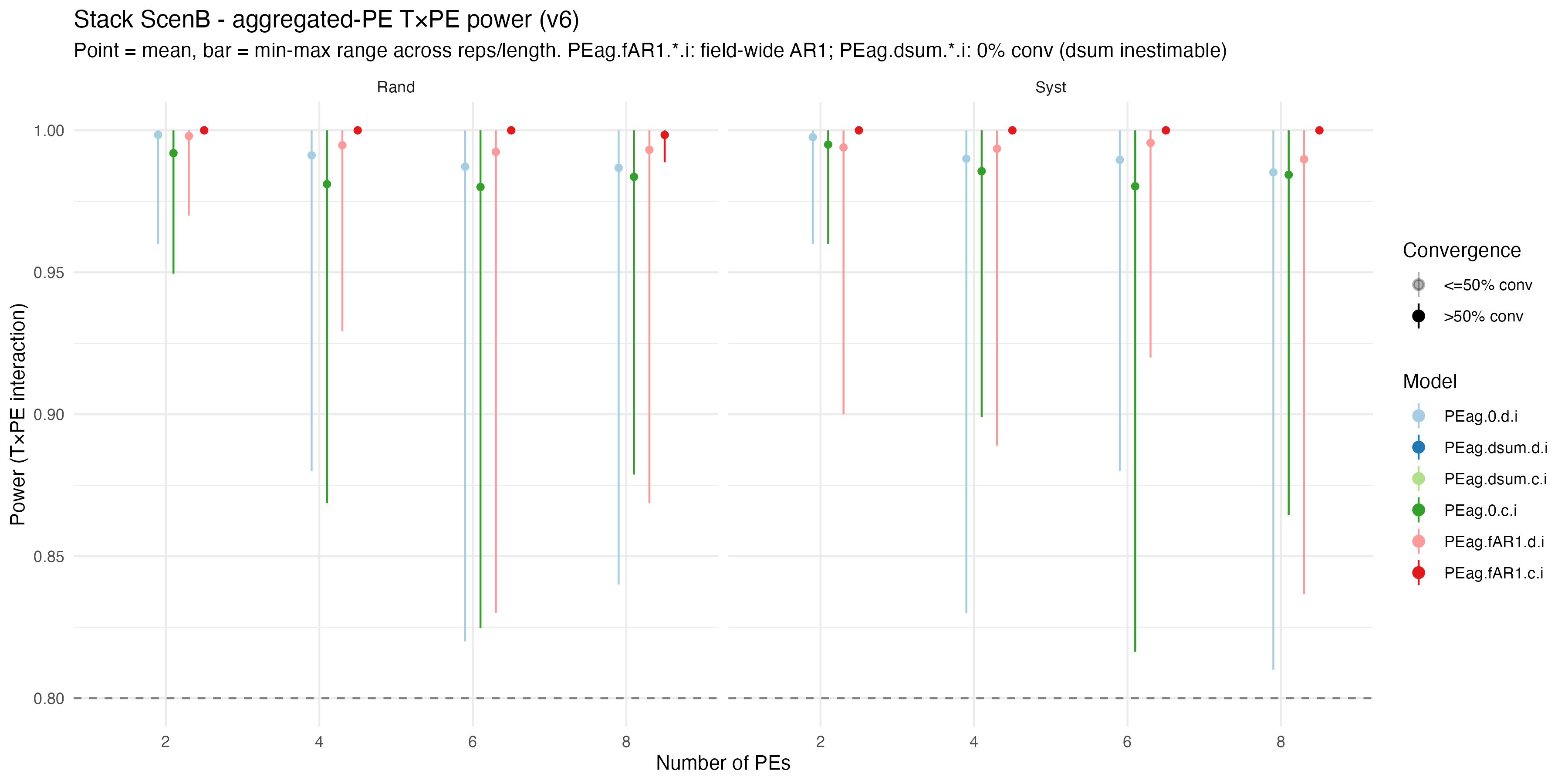}
    \caption{Scenario B (stacked trials): \TPE interaction detection power for the aggregated-PE models.}
    \label{fig:scenB-agg-gxt}
\end{figure}

Table~\ref{tab:power-check} compares predicted and observed power at smaller effects (Section~\ref{sec:detect}), and Figure~\ref{fig:typeI} shows the Type I error of Table~\ref{tab:typeI} by replication. Figures~\ref{fig:strip-power}--\ref{fig:stack-rad} show power at $\beta = 6$ and estimation accuracy (RAD) for the Scenario~A main grid, and Figures~\ref{fig:scenB-rate}--\ref{fig:scenB-agg-gxt} show per-PE rate and \TPE interaction power for Scenario~B.

\begin{table}[H]
\centering
\caption{Predicted and observed power (\%) of the rate test at two smaller response coefficients, at the representative design points (long strip $L = 50$, $r = 4$; stacked $L = 20$, $r = 6$), main grid (centre pixel column). Predicted power uses the error covariance of the design cell (Equation~\ref{eq:mdr}); observed power is from about 300 simulated trials per entry on the same field. Type I is the null rejection rate in the same design cell ($M = 100$). $\beta = 1.5$ and 3 correspond to 0--120\,kg\,N gains of 7.2 and 14.4\% of mean yield.}
\label{tab:power-check}
\small
\begin{tabular}{lllrrrrr}
\toprule
 & & & & \multicolumn{2}{c}{$\beta = 1.5$} & \multicolumn{2}{c}{$\beta = 3$} \\
\cmidrule(lr){5-6}\cmidrule(lr){7-8}
Layout & Allocation & Model & Type I & Predicted & Observed & Predicted & Observed \\
\midrule
Long strip & Randomised & Ag.0        & 7.0 & 62.4 & 66.0 & 99.6 & 98.7 \\
           &            & Ag.AR1      & 4.1 & 94.0 & 92.7 & 100.0 & 100.0 \\
           &            & Px.AR1.plot & 0.0 & 73.0 & 18.4 & 99.9 & 73.7 \\
           & Systematic & Ag.0        & 9.0 & 55.1 & 70.3 & 98.8 & 98.3 \\
           &            & Ag.AR1      & 4.0 & 95.5 & 94.3 & 100.0 & 100.0 \\
           &            & Px.AR1.plot & 1.0 & 57.4 & 24.4 & 99.1 & 75.0 \\
Stacked    & Randomised & Ag.0        & 3.0 & 65.3 & 64.7 & 99.8 & 99.7 \\
           &            & Ag.AR1      & 6.0 & 91.1 & 91.6 & 100.0 & 100.0 \\
           &            & Px.AR1.plot & 7.0 & 51.7 & 55.3 & 98.6 & 95.0 \\
           & Systematic & Ag.0        & 13.0 & 44.5 & 74.0 & 96.6 & 99.7 \\
           &            & Ag.AR1      & 8.0 & 79.1 & 73.4 & 100.0 & 97.3 \\
           &            & Px.AR1.plot & 4.0 & 53.7 & 18.3 & 98.9 & 61.3 \\
\bottomrule
\end{tabular}
\end{table}

At $\beta = 6$, the valid aggregated models reached $\geq 99\%$ power from $r = 3$ on long strips (Ag.0 99.2\%, Ag.AR1 99.4\%) and from $r = 4$ on stacked layouts ($\geq 98.8\%$), whereas the centre-column pixel-level spatial models without a plot term needed more replication on long strips (Px.AR1 93.4\% at $r = 3$, 98.2\% at $r = 4$; Px.AR1w 80.7\% at $r = 3$, 89.9\% at $r = 4$, 99.0\% at $r = 6$). On stacked layouts at $r = 2$ power was 87.6\% (Ag.0), 72.5\% (Ag.AR1) and 86.0\% (Px.AR1). Allocation made at most 0.6 percentage points of difference to power on long strips at $r = 3$. Median RAD was lowest for Ag.AR1 on both layouts (long strip: 4.5\% for $\hat\Delta_{60}$ and 2.2\% for $\hat\Delta_{120}$ under randomisation, against 7.7\% and 3.9\% for Ag.0 and 7.2\% and 3.5\% for Px.AR1; stacked: 5.5\% and 2.7\%, against 8.2\% and 3.9\% for Ag.0 and 11.9\% and 6.2\% for Px.AR1).

\section{PE-aware models: convergence, residual structure and geometry}
\label{app:pe}

\begin{figure}[H]
    \centering
    \includegraphics[width=\textwidth]{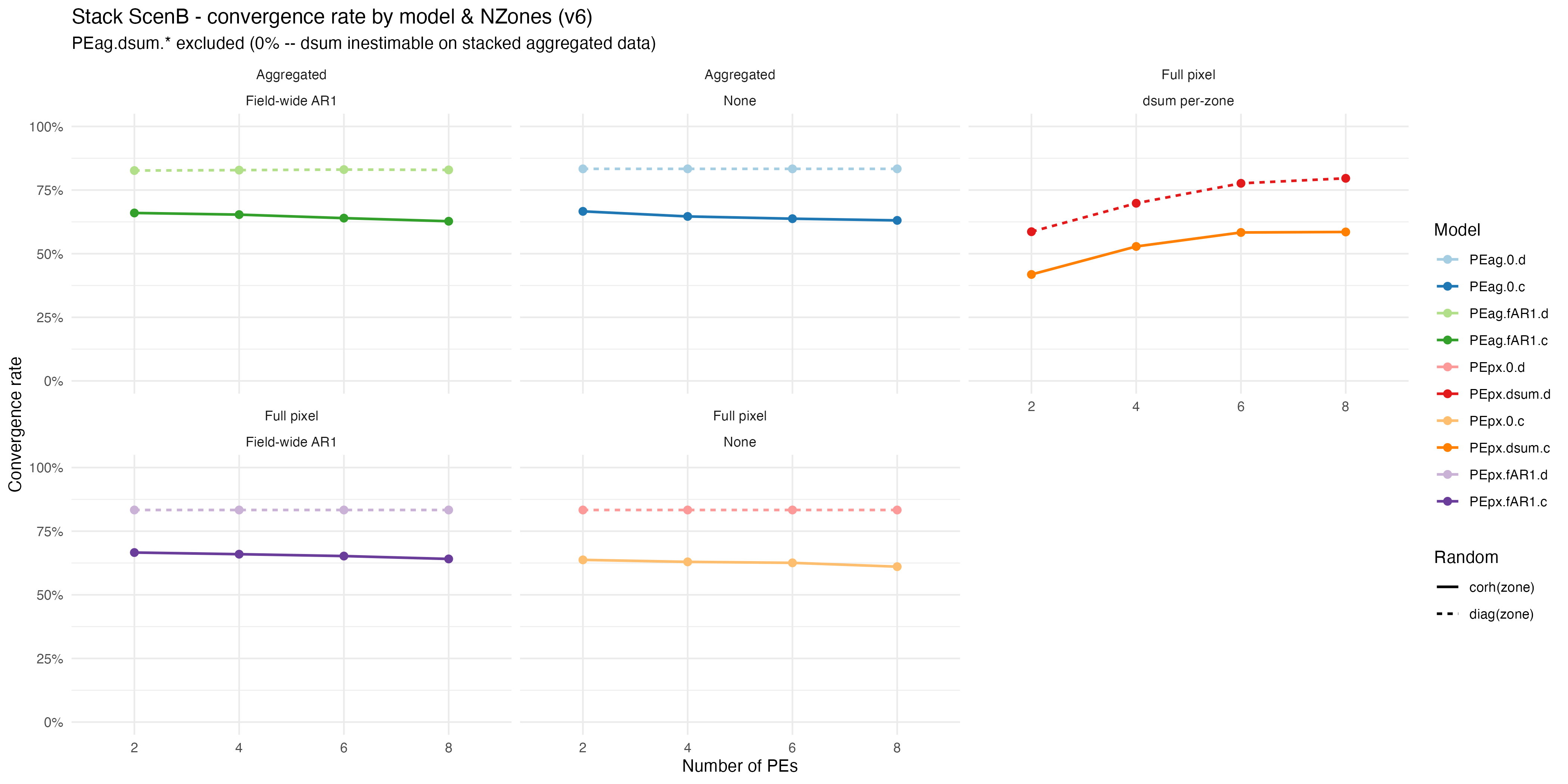}
    \caption{Convergence rates for Scenario B PE-aware models on stacked trials as a function of number of PEs.}
    \label{fig:convergence}
\end{figure}

\paragraph{Convergence.} The dsum per-PE residual models PEag.dsum.d and PEag.dsum.c achieved 0\% convergence on stacked aggregated data: after within-PE aggregation there is only one observation per PE-strip cell, rendering the per-PE AR1 process singular (Table~\ref{tab:convergence}; Figure~\ref{fig:convergence}). PEag.fAR1.d converged in 82.9\% of fits and PEag.fAR1.c in 64.5\%, comparable to the pixel equivalents (PEpx.fAR1.d 83.3\%, PEpx.fAR1.c 65.5\%). In the main grid, combining convergence, validity (3.8--5.5\%), per-PE power (97.7--99.2\%), \TPE power (99.2--99.6\%) and per-PE RAD (13.1\% at the representative design point, against 15.4--17.9\% for the pixel-level models), PEag.fAR1.d was the best centre-column PE-aware model on stacked trials; with all three pixel columns the pixel PE model was more precise (Section~\ref{sec:pe}).

\begin{table}[H]
\centering
\caption{Overall convergence rates for Scenario B PE-aware models on stacked trials (main grid).}
\label{tab:convergence}
\small
\begin{tabular}{lllrr}
\toprule
Model & Data & Residual & N iterations & Conv.\ rate \\
\midrule
PEag.0.d & Aggregated & None            & 24,000 & 83.3\% \\
PEag.dsum.d & Aggregated & dsum per-zone   & 24,000 &  0.0\% \\
PEag.dsum.c & Aggregated & dsum per-zone   & 24,000 &  0.0\% \\
PEag.0.c & Aggregated & None            & 24,000 & 64.5\% \\
PEag.fAR1.d & Aggregated & Field-wide AR1  & 24,000 & 82.9\% \\
PEag.fAR1.c & Aggregated & Field-wide AR1  & 24,000 & 64.5\% \\
PEpx.0.d & Pixel & None            & 24,000 & 83.3\% \\
PEpx.dsum.d & Pixel & dsum per-zone   & 24,000 & 71.4\% \\
PEpx.0.c & Pixel & None            & 24,000 & 62.6\% \\
PEpx.dsum.c & Pixel & dsum per-zone   & 24,000 & 52.9\% \\
PEpx.fAR1.d & Pixel & Field-wide AR1  & 24,000 & 83.3\% \\
PEpx.fAR1.c & Pixel & Field-wide AR1  & 24,000 & 65.5\% \\
\bottomrule
\end{tabular}
\end{table}

\paragraph{diag versus corh.} Across every diag/corh comparison on stacked trials (PEag.0, PEpx.0, PEpx.dsum, PEpx.fAR1, PEag.fAR1), diag(zone) converged 18--21 percentage points more often than the matched corh(zone) variant. For PEag.0, PEpx.0, PEpx.fAR1 and PEag.fAR1 the advantage held at every PE count: diag(zone) remained at $\approx 83\%$ from $K = 2$ to $K = 8$, while corh(zone) declined from $\approx 66\%$ to $\approx 62$--64\%. PEpx.dsum was the exception, with convergence for both structures rising with PE count (diag: 59\% to 80\%; corh: 42\% to 59\%), reflecting identifiability of the per-PE residual rather than estimation of between-PE correlation. When it converged, corh(zone) had higher power by 0.5--2.1 percentage points for PEag.fAR1, 2.7--4.2 for PEpx.fAR1 and 6.4--7.0 for PEpx.dsum, and essentially equal power for PEag.0 and PEpx.0. Because its convergence advantage is large and consistent and its power cost modest, diag(zone) is the recommended default.

\paragraph{Two-dimensional per-PE residual.} Every dsum model in the main grid specifies \texttt{dsum(\allowbreak id(range):\allowbreak ar1(row)\,|\allowbreak\,zone)}, which treats adjacent strips as independent. The two-dimensional counterpart \texttt{dsum(\allowbreak ar1(range):\allowbreak ar1(row)\,|\,zone)} (PEpx.dsum2D.d/c), at the representative design point ($L = 50$, $r = 6$, strips), combined Type I error of 1.0--2.5\% with per-PE power of 95.5--100\% across $K = 2$--10, against 94.5--97.6\% for PEpx.fAR1.d and 85.4--98.5\% for the one-dimensional dsum. Its median per-PE RAD (10.3\%) was lower than that of the one-dimensional form (11.3\%) and similar to that of the field-wide model (10.0\%), and it converged in 99.0\% of fits against 100\% for PEpx.fAR1.d.

\begin{figure}[H]
    \centering
    \includegraphics[width=\textwidth]{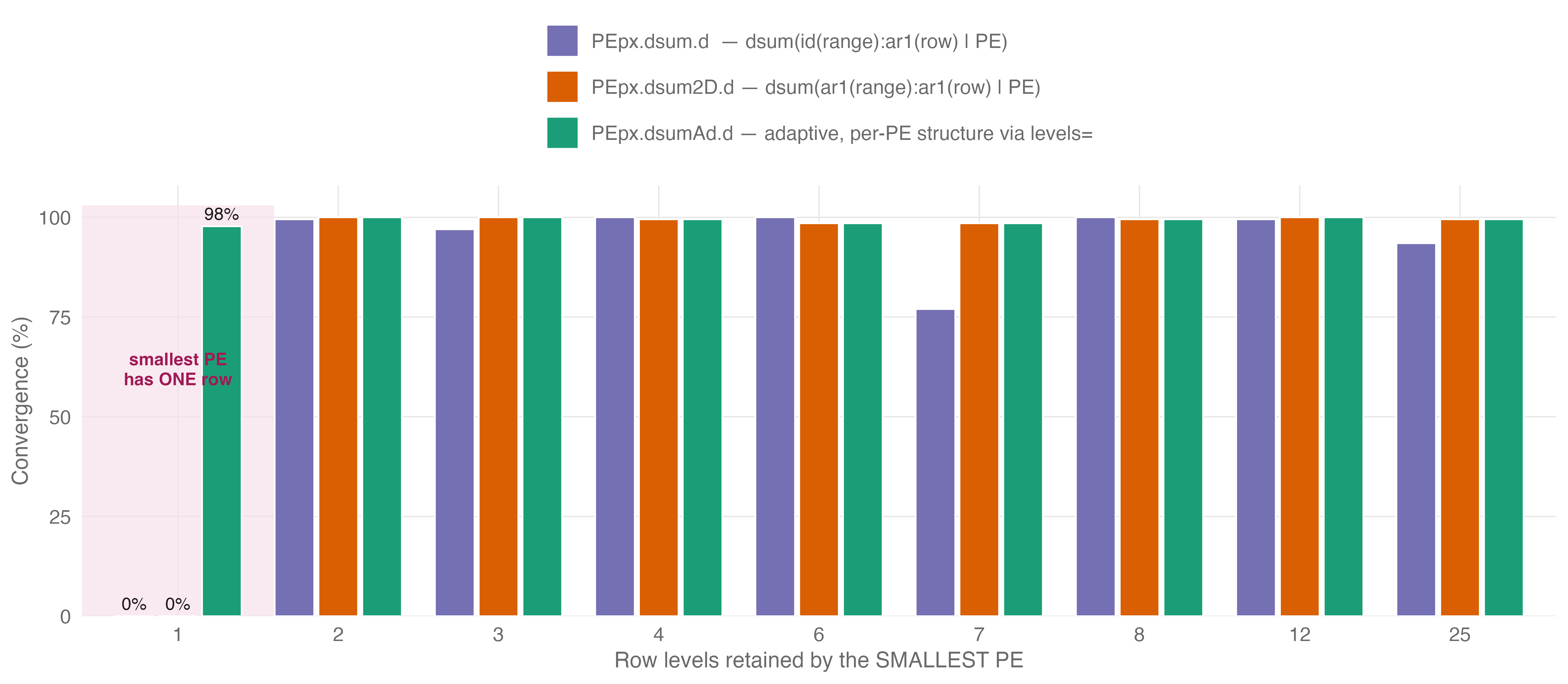}
    \caption{Convergence of three per-PE residual specifications as a function of the number of row levels retained by the smallest PE ($M = 100$ per cell; $n = 400$ at one row).}
    \label{fig:pe-rowthreshold}
\end{figure}

\paragraph{PE geometry.} The per-PE dsum models' convergence is governed by the number of distinct row levels retained by the smallest PE (Figure~\ref{fig:pe-rowthreshold}): at one row both the one- and two-dimensional dsum structures fail (0.0\%), at two rows convergence is 98.6\%, and from three rows upward 95--96\% irrespective of size inequality. Selecting each PE's structure from its own geometry (PEpx.dsumAd.d/c) recovers 97.8\% convergence in the one-row cells, with per-PE power of 98.9\% and Type I error of 3.2\%; where no PE is degenerate it is numerically identical to the fixed two-dimensional form.

\end{document}